\documentstyle[preprint,aps,floats]{revtex}
\tightenlines
\input psfig.tex
\begin{document}
%\rightline{DAMTP-1999-50}
\def\ba{\begin{eqnarray}}
\def\ea{\end{eqnarray}}
\def\be{\begin{equation}}
\def\ee{\end{equation}}
\def\tr{{\rm tr}}
\newcommand{\labfig}[1] {\label{fig:#1}}

\font\boldsym=cmmib12\skewchar\boldsym='177
\textfont9=\boldsym \scriptfont9=\boldsym \scriptscriptfont9=\boldsym

\mathchardef\xib="0918

\title{Characterising the
Primordial Cosmic Perturbations\\
 Using MAP
and PLANCK}

\author{Martin Bucher,\thanks{E-mail: M.A.Bucher@damtp.cam.ac.uk} %
Kavilan Moodley\thanks{E-mail: K.Moodley@damtp.cam.ac.uk} %
and Neil Turok\thanks{E-mail: N.G.Turok@damtp.cam.ac.uk}\\
DAMTP, Centre for Mathematical Sciences, University of Cambridge\\
Wilberforce Road, Cambridge CB3 0WA, United Kingdom}

\date{July 2000}

\maketitle

\begin{abstract}
The most general homogeneous and isotropic 
statistical ensemble of linear 
scalar perturbations which are regular at early times, 
in a universe with only
photons, baryons, neutrinos, and a cold dark matter (CDM)
component, is described by a 5x5 symmetric
matrix-valued generalization of the power spectrum. This description is
complete if the perturbations are 
Gaussian, and even in the non-Gaussian case 
describes all observables quadratic in the small perturbations.
The matrix
valued power spectrum describes the auto-
and cross-correlations of the adiabatic, baryon isocurvature,
CDM isocurvature, neutrino density isocurvature, and 
neutrino velocity isocurvature modes. In this paper we
examine the prospects for constraining or discovering 
isocurvature modes using forthcoming MAP and 
PLANCK measurements of the cosmic microwave background (CMB) anisotropy.
We also consider the degradation in estimates of the cosmological
parameters resulting from the inclusion of these modes. 
In the case of MAP measurements of the temperature 
alone,
the degradation is drastic.
When 
isocurvature modes 
are admitted,
uncertainties in the amplitudes of the mode auto-- and 
cross--correlations, and in the cosmological parameters,
become of order one. With the inclusion of 
polarisation (at an optimistic sensitivity) 
the situation improves for the cosmological parameters but the 
isocurvature modes are still only weakly constrained. 
Measurements with PLANCK's estimated errors are far more constraining,
especially so with the inclusion of polarisation.
If PLANCK operates as planned the 
amplitudes of isocurvature modes will be constrained to less than ten 
per cent of the adiabatic mode and
simultaneously 
key cosmological parameters will be estimated to a few per cent or better.
\end{abstract}

\section{Introduction}

Because the physics by which cosmological perturbations 
imprint themselves on the cosmic microwave backgroun (CMB)  sky
is very nearly linear, CMB observations 
offer a clean and comparatively direct probe of the nature
of the primordial perturbations. Starting with COBE, which 
established a normalization
for the perturbation amplitude on large scales\cite{cobe}.
The dramatic recent results of 
TOCO, BOOMerANG and MAXIMA \cite{toco,boom,maxima} 
indicating a first
Doppler peak, and limiting the amplitude of
the  second, are beginning to severely constrain 
cosmological models.
With the new forthcoming data from
the MAP satellite \cite{mapp} to be launched in Spring 2001,
and from the PLANCK satellite,\cite{plnck} 
to be launched in 2007, the situation
will greatly improve, and it is hoped that one will be
able to determine a host of cosmological parameters 
with great precision from the CMB alone.\cite{paramest}

One of the key assumptions underlying this program is
that that the primordial
fluctuations were adiabatic, that is the relative abundances of 
different particle species were unperturbed from their
thermal equilibrium values. This assumption has the great merit of
simplicity and is even justified in many specific models of the
origin of the fluctuations. 
But given its centrality in
inferences based on the CMB 
anisotropy, it seems worthwhile to attempt to
check the adiabaticity assumption using CMB data. 

From a theoretical point of view, cosmic
inflation is the leading contender as an explanation for the
origin of the perturbations. Inflation 
provides an elegant mechanism for simultaneously explaining 
the large scale structure of the Universe and the 
perturbations \cite{infpert,liddle}. In the simplest inflationary
models the latter are indeed of the simplest form, namely
adiabatic, Gaussian and with
a nearly scale invariant power spectrum. 
But it is important to note that adiabaticity
is not so much a consequence of inflation 
as of the assumption that no additional information beyond the
overall
density perturbation survived the inflationary era. If fields exist which
were perturbed during inflation, and their excitations survived or decayed
in a non-adiabatic manner, then isocurvature perturbations would have 
generically been concomitantly
produced. Many inflationary models have been constructed in 
which this occurs.
Given that there is no currently compelling 
 model, and that most realisations of inflation
within unified theories invoke a great number of additional
fields, 
it seems premature to associate the prediction of adiabaticity
with inflation.
\cite{nongauss}

This situation motivates a more phenomenological
approach to analyzing the new data that contemplates a 
wider range of possibilities for the nature of the 
primordial fluctuations, and seeks to infer 
{\it from the data} what limits may be set on non-adiabatic
perturbations. In a previous paper we examined 
the most general primordial
perturbation possible in a cosmological model with no new
physics---with only baryonic matter (and its associated 
electrons), photons, neutrinos, and a cold dark matter 
component, which is regular at early times.\cite{bmt} {\it Primordial} here refers to the assumption that
the perturbations were generated at a very early time, well before
recombination (at $z\approx 1100$), so that any singular
(i.e., decaying) modes that may have been produced 
had ample opportunity to decay away before leaving an 
imprint on the CMB. In this work we found five regular
modes: an adiabatic growing mode, a baryon isocurvature
mode, a CDM isocurvature mode, a neutrino density
isocurvature mode, and a neutrino velocity isocurvature
mode. The adiabatic growing mode assumes a common equation
of state, spatially uniform everywhere in the 
universe. In the baryon isocurvature mode the ratio
of baryons to photons varies spatially,\cite{pjepa,bondb,cen} 
and similarly in the CDM isocurvature mode the ratio of CDM to
photons varies spatially.\cite{bond} For the neutrino isocurvature
modes, perturbations in the neutrino energy and momentum
densities are balanced by opposing perturbations in
the photon-radiation component, so that at early
times the total stress-energy perturbation vanishes.
At later times, however, the differences in how
photons and neutrinos evolve lead to perturbations
in the total stress-energy, which generate perturbations
in the gravitational potentials, which in turn cause
the baryons and CDM to cluster. Upon entering the 
horizon, the neutrinos free stream behaving as a 
collisionless fluid, while the photons behave like
a perfect fluid because of the strong Thompson scattering
off free electrons. 
The neutrino isocurvature modes, discussed in detail
in ref. \cite{bmt}, are implicit in the work of Rebhan and
Schwarz\cite{nnewb} and of Challinor and Lasenby\cite{nnewc};
however, these authors do not investigate their implications.
If Gaussian perturbations produced by a spatially homogeneous
and isotropic random process are assumed, the most
general perturbation of these five modes is 
completely described by the 5x5 symmetric correlation matrix 
\be
P_{ij}({\bf k})=\langle A_i({\bf k})~A_j(-{\bf k})\rangle 
\ee
where $(i,j=1,\ldots ,5)$ labels the modes and 
$A_i({\bf k})$ indicates the strength of the $i$th 
mode with wavenumber ${\bf k}.$ 
This  generalizes the usual scalar power 
spectrum. In the case of non-Gaussian perturbations,
the above matrix suffices to determine the 
expectation values for all perturbations 
quadratic in the small perturbation (this includes
the CMB power spectrum) if one assumes that the 
linearized theory is valid, which is a good
approximation. 
The possibility of ascribing 
cosmological perturbations entirely to isocurvature
modes, either to the baryon isocurvature mode 
or to the CDM isocurvature mode,
has previously been considered, and these possibilities were
found inconsistent with the existing observational 
data.\cite{stompor}--\cite{nnewa}
However, apart from the work of Enqvist and
Kurki-Suonio\cite{enqvist} and of Pierpaoli, Garc\'ia-Bellido, 
and Brogani\cite{pierpaoli}
little effort has been devoted to the problem of 
detecting or constraining admixtures of isocurvature
modes observationally. Moreover, when one admits the
possibility of more than one mode being excited, 
the possibility arises of correlations between the several
modes. These are characterized by the off-diagonal elements
of $P_{ij}({\bf k}).$ Linde and Mukhanov\cite{lindeb},
Langlois,\cite{langl}
and Langlois and Riazuelo\cite{langlr},
drawing on the work of Polarski\cite{polarski}
on double inflation, 
investigated an inflationary model with two scalar fields
exciting both the adiabatic and baryon isocurvature modes
in varying proportions and with various degrees of correlation
between these modes. In an inflationary model with five
or more scalar fields (or a single field with the same
number of real components), it is generically possible to
realize the most general $P_{ij}({\bf k})$ of the form
discussed above.

The outline of the paper is as follows. Section II discusses
the statistical techniques employed to interpret CMB 
measurements and reviews the properties of the five
regular perturbation modes we allow. Section III gives
numerical results in the form of eigenvectors and eigenvalues
of the `Fisher matrix', corresponding to the MAP and 
PLANCK measurements. The Tables provided allow one to
calculate the uncertainty in any cosmological 
parameter or function thereof, as well as the
amplitudes of the primordial perturbation modes.
Section IV discusses the significance of the results, 
and our main conclusion, which is that a high precision
measurement of the cosmic polarisation will be essential
to an accurate determination of the cosmological
parameters and the primordial perturbations.

We should mention three limitations of our analysis. 
Our main goal is to explore the effect of relaxing the assumption
of adiabatic perturbations. Therefore we 
ignore tensor modes, which would complicate the discussion.
But many inflationary models do predict tensor modes and 
it is important they be included in any complete analysis. 
The second caveat is that we are only considering very sub-dominant 
isocurvature perturbations here - our analysis is perturbative around
a standard $\Lambda$ CDM model. Finally 
we do not vary the spectral indices (or the shape of the 
spectra) for the isocurvature modes. We consider only
`scale invariant' isocurvature perturbations, as defined below.
In inflationary models,
isocurvature perturbations (like adiabatic ones) would in general
have adjustable power spectra and again that would complicate the
discussion.

\section{The General Primordial Cosmic Perturbation}

Statistical analysis of observational data, be it based
on `classical' or `frequentist' statistics or on Bayesian
statistics, ultimately reduces to considering 
relative likelihoods of competing theoretical descriptions 
given the observed data. Because the observational data is not 
yet available, we must assume a particular underlying theoretical model
for computing expectations for relative likelihoods.
We assume a statistical
model with independent Gaussian distributions for the individual 
moments of the CMB multipole expansion with variance
\be 
\langle \vert a_{\ell m}\vert ^2\rangle = 
c_{\ell }+\sigma ^2_{n, \ell }
\label{booa}
\ee
where $c_{\ell }$ is the variance of the underlying 
cosmological signal and $\sigma ^2_{n, \ell }$ is the 
variance resulting from detector noise.
It follows that the expectation value of the logarithm of the 
relative likelihood of model B relative to model A under the 
assumption that the data was produced according the distribution
from model A is given by the formula
\ba
\left\langle \log \left[ {p_B\over p_A}\right]\right\rangle _A
&=&\left\langle \log \left[ {p(\{ a_{\ell m}\}|B)
\over p(\{ a_{\ell m}\}|A)}\right] \right\rangle _A\nonumber\\
&=&{f_{sky}\over 2}\sum _{l=2}^{\ell _{max}}(2\ell +1)
\left\{1-
{c_{\ell ,A}+\sigma ^2_{n, \ell }\over c_{\ell ,B}+\sigma ^2_{n, \ell }}
+\log \left[
{c_{\ell ,A}+\sigma ^2_{n, \ell }\over c_{\ell ,B}+\sigma ^2_{n, \ell }}
\right]
\right\}
\label{boob}
\ea
where $\{ a_{\ell m}\}$ is the observed data, $c_{\ell ,A}$ and $c_{\ell ,B}$
are the variances of the cosmological signal predicted by models
$A$ and $B,$ respectively, and $\sigma ^2_{n, \ell }$ is the Gaussian 
detector noise for each multipole of order $\ell .$ $f_{sky}$ indicates
the fraction of the sky remaining after the galaxy cut. 

We consider cosmological models where $c_\ell $ depends on
a number of continuous parameters $\alpha _1,\ldots ,\alpha _N.$
With relatively little detector noise and with changes in 
the parameters affecting a large number of multipoles, 
in the region of interest, where the likehood relative to the 
reference model is comparable, the following quadratic approximation 
is justified
\ba &&\left\langle \log \left[ 
{ p(\alpha _1,\ldots ,\alpha _N)
\over p(\alpha _1^{(0)},\ldots ,\alpha _N^{(0)}) }
\right]\right\rangle _{(0)}\nonumber\\
&&\quad  = -{f_{sky}\over 2}
\sum _{i,j=1}^N
\sum _{l=2}^{\ell _{max}}
(2\ell +1)\times {1\over (c_{\ell ,B}+\sigma ^2_{n, \ell })^2}
{\partial c_\ell \over \partial \alpha _i}
{\partial c_\ell \over \partial \alpha _j}
{(\alpha _i-\alpha _i^{(0)})
(\alpha _j-\alpha _j^{(0)})\over 2}.
\label{booc}
\ea
We write the right hand side 
as $-{1\over 2} F_{ij} (\alpha _i-\alpha _i^{(0)} )
(\alpha _j-\alpha _j^{(0)}) $.
In the case where the variations in the logarithm are
described by a quadratic form, 
the matrix $F_{ij}$ is equivalent
to the Fisher matrix, and in the sequel we shall refer to 
it in this way even though the
terminology is not completely accurate. 
The quadratic approximation may break down when considering
variations primarily affecting very low-$\ell $ moments, where
a Gaussian approximation to a $\chi ^2$-distribution of 
low order is inaccurate even near where it is peaked.  

When polarization is included, eqn. (\ref{booa}) is modified to
become
\be
M_\ell =\pmatrix{
\vert \langle a_{\ell m}~ a_{\ell m} \rangle \vert &
\vert \langle a_{\ell m}~ b_{\ell m} \rangle \vert \cr 
\vert \langle b_{\ell m}~ a_{\ell m} \rangle \vert &
\vert \langle b_{\ell m}~ b_{\ell m} \rangle \vert \cr 
}=
\pmatrix{
c_{\ell , T}+\sigma ^2_{n\ell , T}&
c_{\ell , C}\cr 
c_{\ell , C}&
c_{\ell , P}+\sigma ^2_{n\ell , P}\cr 
}
\ee
and (\ref{boob}) is modified to
\ba
\left\langle \log \left[ {p_B\over p_A}\right]\right\rangle _A
&=&\left\langle \log \left[ {p(\{ a_{\ell m}\},~
\{ b_{\ell m}\}|B)
\over p(\{ a_{\ell m}\},~\{ b_{\ell m}\}|
A)}\right] \right\rangle _A\nonumber\\
&=&{f_{sky}\over 2}\sum _{l=2}^{\ell _{max}}(2\ell +1)
\Biggl[ 
\tr \biggl\{I -M_{\ell A}{M_{\ell B}}^{-1}\biggr\}
+\ln \left\{ \det \biggl( M_{\ell A}{M_{\ell B}}^{-1}\biggr)\right\} 
\Biggr] 
\ea
The second derivative of the above may be expressed as
\ba
\delta ^2&&\left\langle \log \left[ {p_B\over p_A}\right]\right\rangle _A 
=f_{sky}\sum _{l=2}^{\ell _{max}}(2\ell +1)\nonumber\\ 
&&~~~\times \Biggl[
\tr \biggl\{ 
(\delta M_\ell ) {M_{ref, \ell }}^{-1}
(\delta M_\ell ) {M_{ref, \ell }}^{-1}
\biggr\}
+\det \biggl\{ 
(\delta M_\ell ) {M_{ref, \ell }}^{-1}
\biggr\}
-{1\over 2}\tr ^2\biggl\{ 
(\delta M_\ell ) {M_{ref, \ell }}^{-1}
\biggr\}
\Biggr].
\ea

Most studies of parameter estimation using future CMB data assume
adiabatic perturbations from inflation allowing a set of model
parameters, such as
$H_0,$ $\Omega _b,$ $\Omega _\Lambda $ for example, to be varied.
A fiducial, or reference, model is assumed, and the behavior of the 
relative likelihood in the neighborhood of this reference model
is explored.  

We extend this approach by including parameters for 
the strengths of the isocurvature modes and of their cross-correlations, 
both with respect to one another as well as to the adiabatic mode. 
In this case the parametric model becomes as follows
\ba 
c_{\ell }(
\mbox{\boldmath $\beta $\unboldmath }, 
\mbox{\boldmath $\gamma $\unboldmath }, 
\mbox{\boldmath $\delta $\unboldmath }, 
\mbox{\boldmath $\xi $\unboldmath }
)&&=
c_{\ell }^{adia}(\beta _1,\ldots ,\beta _N)
+\sum _{A=1}^4\gamma _A ~
c_{\ell }^{A}(\beta _1,\ldots ,\beta _N)\nonumber\\
&&+\sum _{A=1}^4\delta _A ~
c_{\ell }^{A-adia}(\beta _1,\ldots ,\beta _N)
+\sum _{A,B=1\atop A\ne B}^4~\xi _{AB}~
c_{\ell }^{cross, AB}(\beta _1,\ldots ,\beta _N)
\ea
where the indices $A,$ $B$ label the four nonsingular
isocurvature modes. 
Here the vector \boldmath $\beta ~$\unboldmath  represents 
the usual cosmological parameters. The vector 
\boldmath $\gamma ~$\unboldmath   indicates the 
auto-correlations of the isocurvature modes.
The vector \boldmath $\delta ~$\unboldmath  indicates the 
cross-correlation 
of the isocurvature modes with the adiabatic mode, 
and symmetric off-diagonal elements 
$\xi _{AB}$ indicates the cross correlations of the 
isocurvature modes with each other. 
The components $c_{\ell }^{A}$ are computed with 
a modified version of CMBFAST with the isocurvature
mode excited. The cross correlations are computed
by running CMBFAST first with two modes excited and then subtracting
the results when each mode individually excited,
using $Q(A,B)={1\over 2}[Q(A+B,A+B)-Q(A,A)-Q(B,B)]$ for a quadratic form
$Q$.
{}From the viewpoint of statistical analysis described above,
all components of the combined  vector 
\boldmath
$\alpha =(\beta , \gamma ,\delta ,\xi )$
\unboldmath
stand on equal footing. 
The three components
\boldmath $\gamma $\unboldmath , 
\boldmath $\delta $\unboldmath , 
\boldmath $\xi ~$\unboldmath  are related to the correlation
matrix M given in ref.\cite{bmt} according to
\be 
M=\pmatrix{
1& \delta _1& \delta _2& \delta _3& \delta _4\cr
\delta _1& \gamma_1&   \xi _{1,2}& \xi _{1,3}& \xi _{1,4}\cr
\delta _2& \xi _{2,1}& \gamma _{2}& \xi _{2,3}& \xi _{2,4}\cr
\delta _3& \xi _{3,1}& \xi _{3,2}& \gamma _{3}& \xi _{3,4}\cr
\delta _4& \xi _{4,1}& \xi _{4,2}& \xi _{4,3}& \gamma _{4}\cr
}
\ee
Some free parameters of our model cannot be treated
in the same way as the ones above when the fiducial model 
is purely adiabatic with no isocurvature component excited.
These include the spectral indices for the isocurvature modes.
The difficulty arises because near the fiducial model the 
dependence on these spectral indices vanishes. In this paper
we shall simply fix 
the spectral indices to correspond to `scale invariant'
isocurvature perturbations. For simplicity 
we also choose the cross correlation 
power spectra to be just the geometric mean of the autocorrelation power
spectra of the same two variables. This assumption is 
straightforward to generalise, and we shall do so elsewhere.

\begin{figure}
\centerline{\psfig{file=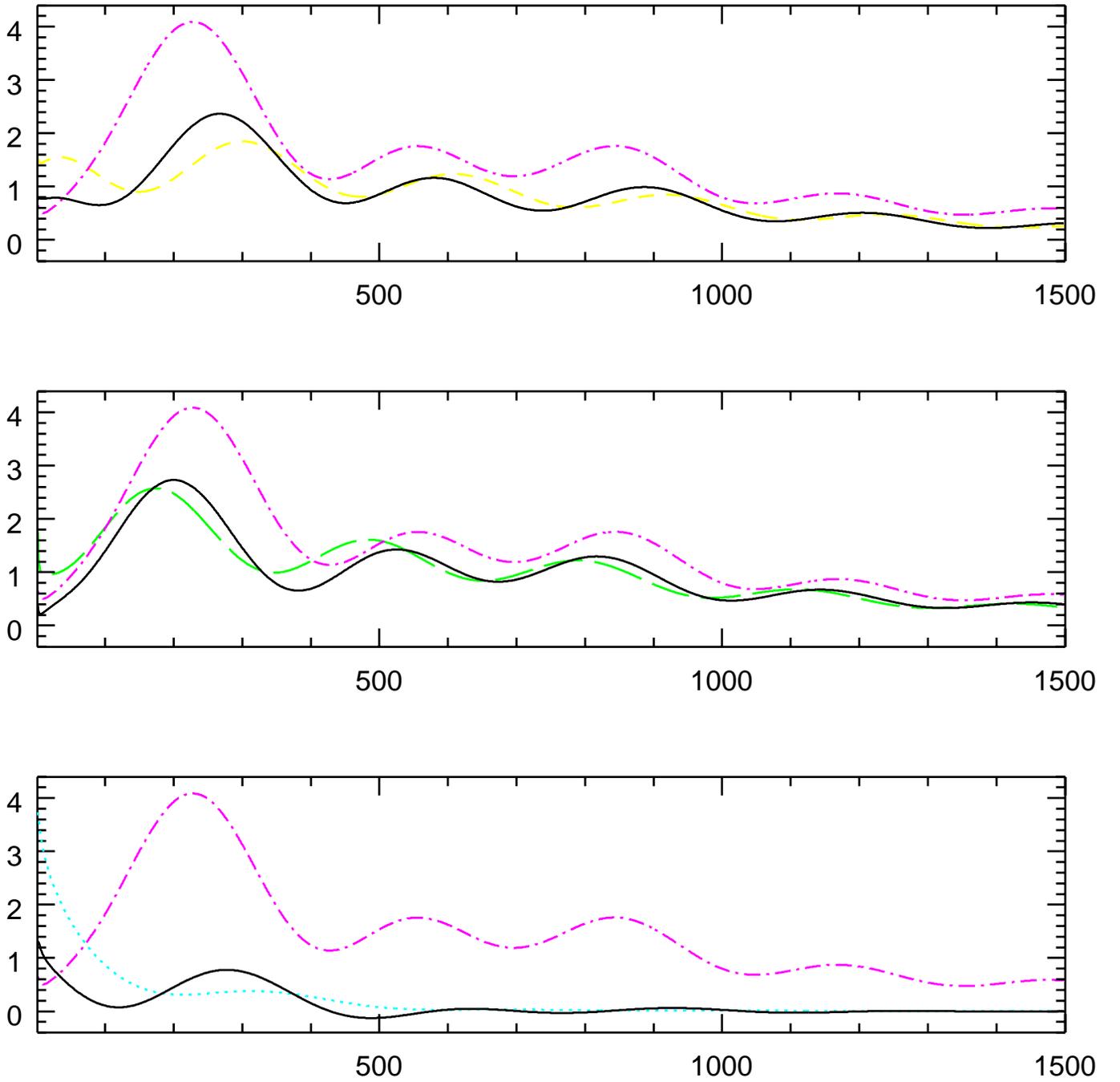,width=8.in}}
\caption{CMB anisotropy power spectra $l(l+1)C_l$ 
are plotted versus $l$.
 The three plots show 
cross-correlation 
power spectra as solid lines:
adiabatic-baryon isocurvature  density (lower), adiabatic-neutrino isocurvature  density
(middle) and adiabatic-neutrino isocurvature  velocity (upper). The relevant
auto-correlation spectra are also shown on each plot as follows:
adiabatic perturbations (dot-dashed, magenta), 
baryon isocurvature (dotted, cyan), neutrino density isocurvature
(short dashed, yellow), and neutrino velocity isocurvature 
(long dashed, green). All are assumed to have 
scale invariant underlying 
power spectra.}
\end{figure}
\bigskip

\begin{figure}
\centerline{\psfig{file=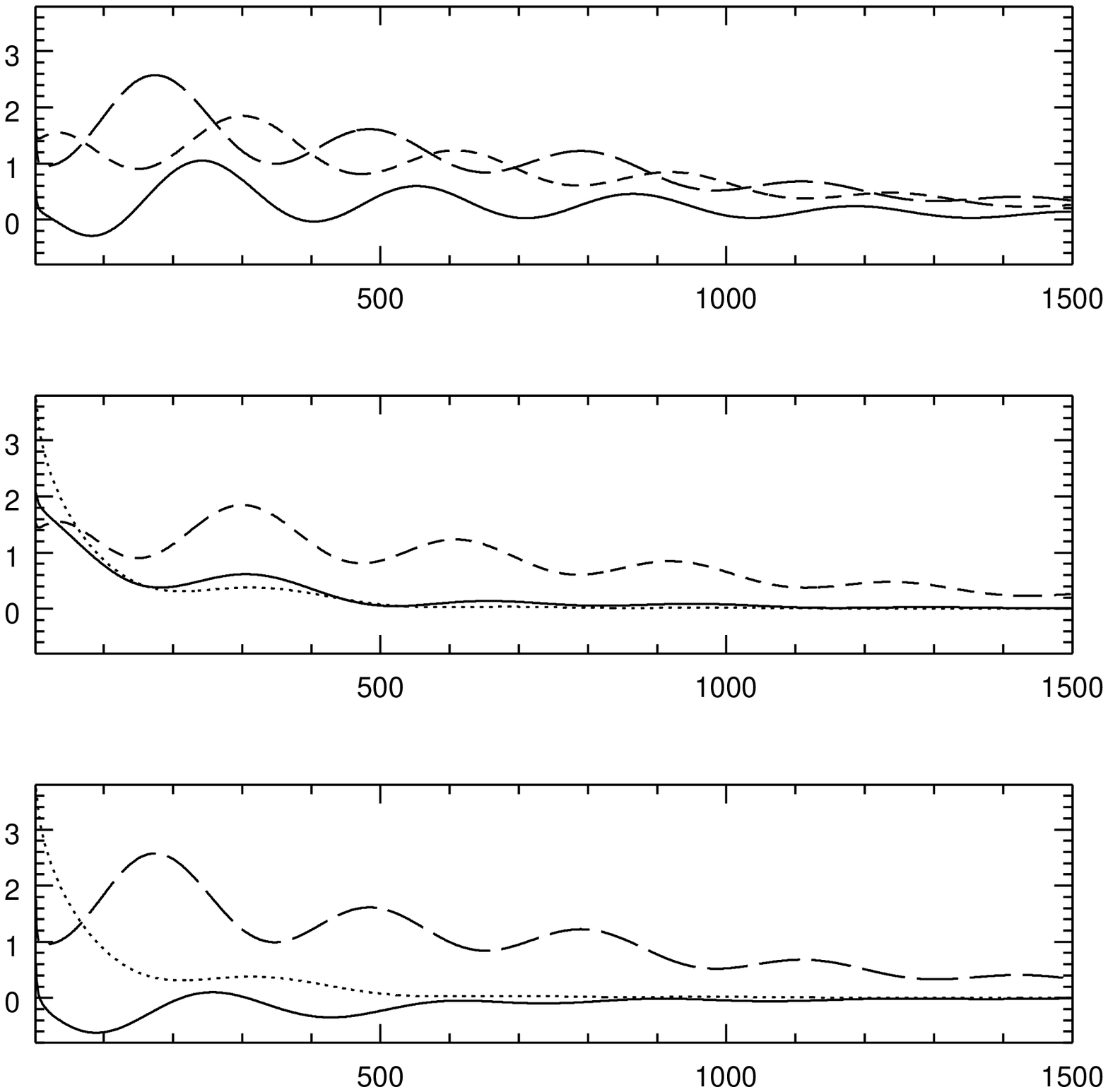,width=8.in}}
\caption{As in Figure 1, but for the baryon isocurvature -neutrino isocurvature  velocity 
correlation (lower),  baryon isocurvature-neutrino isocurvature  density correlation (middle)
and neutrino isocurvature  density-neutrino isocurvature  velocity correlation (upper).
On each plot the
 cross-correlation power spectrum is shown as a solid black line, 
with the corresponding 
auto-correlation power spectra denoted as in Figure 1.}
\end{figure}

Figures 1 and 2 
show the $C_l$ power spectra for the modes we allow here. 
We choose our fiducial cosmological model to be 
$h=0.65,$ $\Omega _b=0.06,$ $\Omega_\Lambda =0.69,$
$\Omega _{cdm}=0.25,$
$n_S=1,$ and a small amount of reionization with an
optical depth to the last scattering surface of $\tau =0.1.$
In Figure 1, $l(l+1) C_l$ from adiabatic scale-invariant perturbations
is plotted against the same quantity for baryon isocurvature,
neutrino isocurvature density and neutrino isocurvature velocity modes.
For the baryon isocurvature mode  a scale free spectrum of 
$\delta(\rho_B/\rho_\gamma)$ is assumed i.e. the variance 
at early times is a
logarithmically divergent integral over wavenumber. 
For neutrino isocurvature density perturbations a scale
free spectrum is assumed for $\delta(\rho_\nu/\rho_\gamma)$
and for neutrino isocurvature velocity  perturbations a scale
free spectrum is assumed for the bulk neutrino velocity $v_\nu$. 
The CMB spectra for the CDM isocurvature mode agree with those
obtained for the baryon isocurvature mode to a fraction of
a percent. Therefore, we do not consider the CDM case separately.
The plots also show the $l(l+1) C_l$ produced from the cross
correlation of the allowed modes.

\vfill\eject

\section{Numerical Results}

We now proceed to our numerical results. Both the
MAP and PLANCK experiments are considered. Following
\cite{wang}, we assume a fractional sky coverage
$f_{sky}=0.8$ after the galaxy cut and ignore
possible foreground contamination. We further
assume that the several channels may be combined,
using for each multipole the linear combination
giving the least variance for the noise. In
other words, we assume that linear combinations
chosen to project out foreground contamination
based on differing spectral properties are not
necessary. Under these assumptions, it follows that
\be 
{1\over \sigma ^2_{n,\ell }}=\sum _c {B_{\ell ,c}^{2}\over
\sigma ^2_c~\theta _{fwhm,c}^2}
\ee
where $c$ indicates the sum over channels,
$B_{\ell ,c}^{2}=\exp [-(0.425~\theta _{fwhm,c}~\ell )^2]$
is the window function assuming a Gaussian beam,
$\theta _{fwhm,c}$ is the full width at half maximum
of the $c$th channel, and $\sigma ^2_c~\theta _{fwhm,c}^2$
is the mean square noise per multipole. The assumption
of no correlations between pixels makes this quantity
independent of $\ell .$ For MAP we assume three channels
with $\theta _{fwhm,c}=0.47^\circ ,$ $0.35^\circ ,$
and $0.21^\circ $ with $\sigma = 17.2\mu K,$ $30\mu K,$ and
$50\mu K,$ respectively, as anticipated after
two years of data.\footnote{$\sigma = 27\mu K,$ $35\mu K,$ and
$35\mu K$ are the corresponding temperature errors for
a $0.3^\circ $ square pixel given in the Wang, Spergel,
and Strauss paper.\cite{wang}}
For Planck, we use the three
lowest frequency channels of the High Frequency
Instrument (HFI) with $\theta _{fwhm,c}=0.18^\circ ,$
$0.13^\circ ,$ and $0.092^\circ $ with $\sigma _c=4.5\mu K,$
$5.5\mu K,$ and $11.8\mu K,$ respectively, as expected
after 12 months of data (which would correspond to
$2.7\mu K,$ $2.4\mu K,$ and $3.6\mu K$ on a $0.3^\circ $
square pixel). 

We also indicate below the result of including polarization.
For the channels for which polarization information is 
available 
(all three MAP channels used here 
and the two higher frequency of the three Planck channels),
we assume that $\sigma _{nP}^2 = 2  \sigma _{nT} ^2.$
This assumption is optimistic for MAP. The experiment has not
been optimised for a polarisation measurement, and 
systematics and foreground effects not included here 
will in all likellihood dominate the polarisation signal.
Nevertheless we include 
calculations making the most optimistic assumption for comparative 
purposes. 

We do not indicate an uncertainty in normalization, because
a natural way to compare relative normalizations between 
spectra of different shapes is lacking. Other authors
indicate the uncertainty in the predicted expectation 
value for the quadrupole. This, however, is a not a very 
useful quantity given the large uncertainty from cosmic
variance in the quadrupole. We instead marginalize over the 
normalization. In the quadratic approximation this is 
equivalent to using the best fit normalization for the 
given values of the parameters. This amounts to replacing the Fisher 
matrix with the reduced Fisher matrix
\be
\hat F_{ij} = F_{ij}-{ F_{i0} F_{j0}\over  F_{00}}.
\ee
where the index 0 labels the parameter describing the overall
normalisation of the power spectrum.

In the following subsections we present tables of the eigenvalues 
and eigenvectors (spectral decomposition) 
of the reduced Fisher matrix $\hat F_{ij}$. 
These contain a wealth of information, which is why we 
include them in this preprint version of the paper. 
They enable one to directly determine (in the Gaussian approximation) 
the uncertainties in any 
cosmological parameters or perturbation amplitudes. They also clearly
illustrate which quantities are and are not accurately determined 
by the measurements. We hope the reader enjoys perusing these tables. 

\subsection{Fiducial Model with No Isocurvature Modes}

Table 1-M-T presents the spectral decomposition 
of the Fisher matrix for adiabatic perturbations in
a spatially flat cosmological model 
with $h=0.65,$ $\Omega _b=0.06,$ $\Omega_\Lambda =0.69,$
$\Omega _{cdm}=0.25,$ 
$n_S=1,$ and a small amount of reionization with an 
optical depth to the last scattering surface of $\tau =0.1.$
Variations in the parameters $H_0,$ $\Omega _\Lambda ,$ 
$\Omega _b,$ $n_S,$ $\tau ,$ and $\Omega _k$ are considered. 
Here $n_S$ is the exponent of the power law for the scalar
power spectrum and the spatial curvature is characterized 
in terms of $\Omega _k=(1-\Omega _\Lambda -\Omega _{cdm}-\Omega _b)$
Except for the last two of these parameters, all variations
are considered fractionally. We give
the eigenvectors of the Fisher matrix and the percentage
error associated with measuring each eigenvector component
(i.e., $100\times \lambda ^{-1/2}).$ This table uses the noise
parameters for the MAP satellite with only the temperature 
anisotropy taken into account. The lower table in Table 1 indicates the 
percentage errors in determining each of the cosmological parameters
parameters---that is, $\sqrt{{(\hat F^{-1})}_{ii}} $ for the 
parameter labeled by $i$.

Tables 1-M-TP
repeat the above analysis but using 
both temperature
and polarization as well as the cross correlation between
the temperature and polarization multipole moments.
Table 1-P-T, and 
1-P-TP are the analogues for the 
PLANCK satellite.

\subsection{Fiducial Adiabatic Model Plus One Isocurvature Mode}

In this subsection we consider models exactly one isocurvature
mode excited in addition to the adiabatic mode. We normalize
the isocurvature modes so that their mean square power 
contributed the the CMB temperature anisotropy summed from 
$\ell =2$ to $\ell =1500$ is equivalent to that of the adiabatic
mode. For one additional mode, two new parameters are added 
in addition to those described in the previous section:
the isocurvature auto-correlation $\langle II\rangle $
and the cross-correlation with the adiabatic mode
$\langle AI\rangle .$ The positive definiteness of the
matrix valued power spectrum requires that 
$\vert \langle AI\rangle \vert \le \langle II\rangle .$

\subsubsection{Baryon Isocurvature Mode}

Tables 2-M-T, 2-M-TP, 
2-P-T and  2-P-TP perform the above analysis
where $I$ indicates the baryon isocurvature mode.  
A scale-free spectrum for the baryon isocurvature mode
is assumed, as defined above, and the cross correlation 
power spectrum is the geometric  mean of the adiabatic
and isocurvature spectra.

\subsubsection{Neutrino Density Isocurvature Mode}

Tables 3-M-T, 3-M-TP,
3-P-T and  3-P-TP perform the above analysis
where $I$ indicates the neutrino density isocurvature mode,
taken to have a scale free spectrum.

\subsubsection{Neutrino Velocity Isocurvature Mode}

Tables 4-M-T, 4-M-TP, 
4-P-T and 4-P-TP perform the above analyis
where $I$ indicates the neutrino velocity isocurvature mode.
Again the velocity mode is taken with a scale free spectrum.

\subsection{Fiducial Adiabatic Model Plus Two Isocurvature Modes}

With two isocurvature modes $I_1$ and $I_2$ there are 
five additional parameters: two auto-correlations
$\langle I_1I_1\rangle $ and $\langle I_1I_2\rangle ,$
and three cross correlations $\langle AI_1\rangle ,$
$\langle AI_2\rangle ,$ and $\langle I_1I_2\rangle .$
Again there is a constraint on these parameters
arising from the requirements of positive definiteness
of the matrix-valued power spectrum.

For Tables 5-M-T, 5-M-TP,
5-P-T and 5-P-TP, 
$I_1$ is the neutrino density isocurvature 
mode and $I_2$ is the baryon isocurvature mode.
For Tables 6-M-T, 6-M-TP, 
6-P-T and 6-P-TP, $I_1$ is the 
neutrino velocity isocurvature mode and $I_2$ is the 
baryon isocurvature mode.
For Tables 7-M-T, 7-M-TP, 7-P-T
and 7-P-TP, $I_1$ is the neutrino density 
isocurvature mode and $I_2$ is the 
neutrino velocity isocurvature mode.

\subsection{Fiducial Adiabatic Model Plus Three Isocurvature Modes}

We now include the  baryon isocurvature and both neutrino
isocurvature modes. There are now three auto-correlations
and six cross-correlations. The results are indicated in Tables 
8-M-T, 8-M-TP, 8-P-T and 8-P-TP.
Given the large uncertainties (of order unity) in all cases except
P-TP, 
the quadratic approximations
employed are no longer accurate. Moreover, the large
ratios between the largest and smallest eigenvalues
make the calculations sensitive to small
errors in the computing the Fisher matrix elements,
since the smallest eigenvalues control the largest errors. 
Nevertheless, the result that the errors in the parameters are
large (of order unity) is reliable. 

\section{Discussion}
\begin{figure}
\centerline{\psfig{file=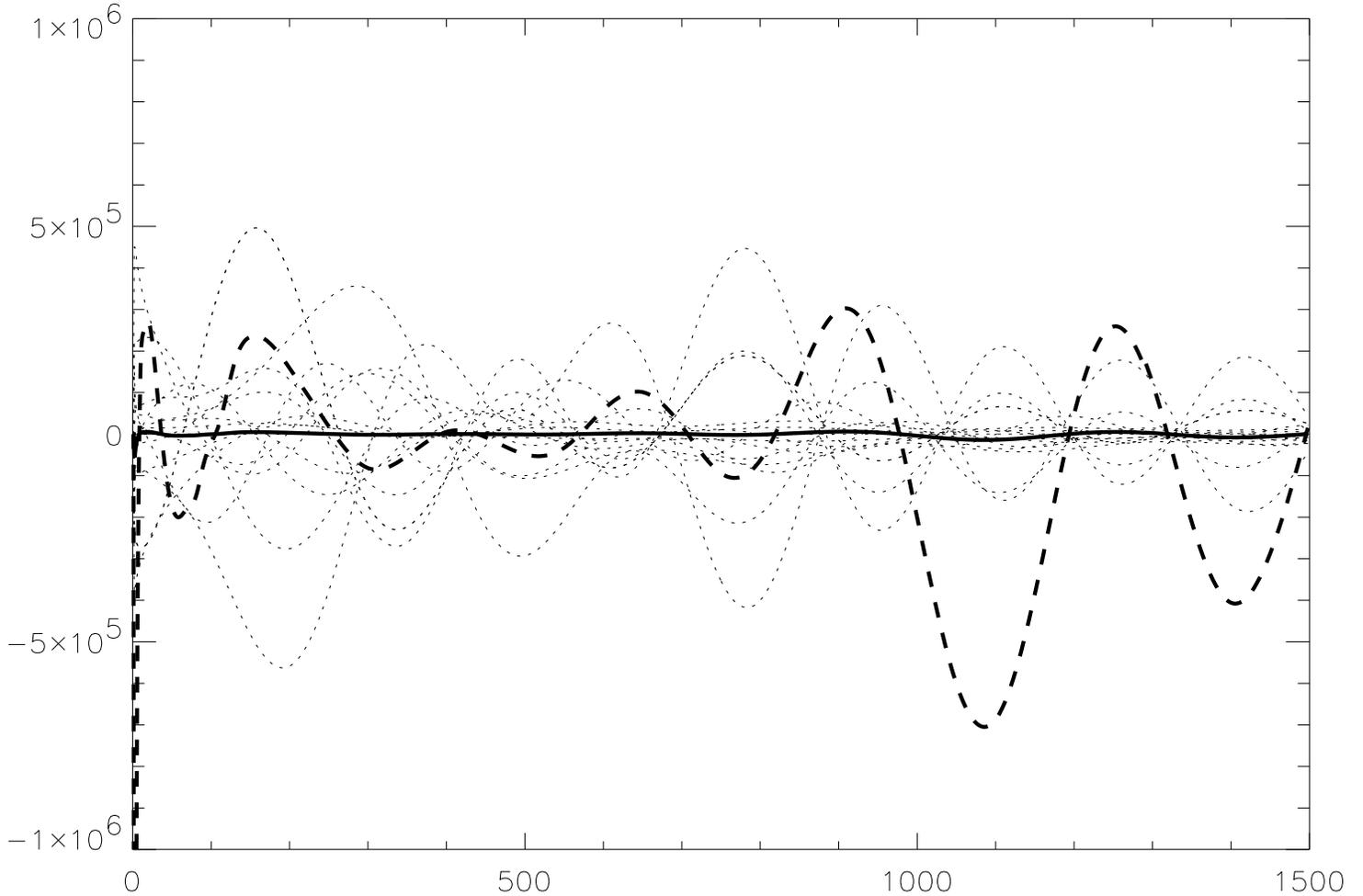,width=8.in}}
\caption{Illustration of the degeneracy problem. Including isocurvature
modes renders the determination of cosmological parameters from
the MAP satellite hazardous. Deviations from the fiducial 
adiabatic model in certain directions in the space of cosmological
parameters and density perturbation amplitudes produce nearly 
degenerate $C_l$ spectra. Here we show the deviations 
 $\delta
l(l+1) C_l$ for each  parameter change multiplied by the appropriate
component of the 
eigenvector of $\hat F_{ij}$ with smallest eigenvalue (given in Table
8-M-TP).
The dotted lines show the individual contributions as a function of $l$,
and the 
solid line the sum after the normalisation has been 
projected out. The heavy dashed line shows the sum multiplied by 50
to render it more visible. The latter deviation is 
not measurable with
MAP, but is with PLANCK, presumably because the latter is more
sensitive to the high $l$ structure.
}
\end{figure}
\bigskip

We have considered the question of to what extent it will be possible to
check
the assumption that the primordial perturbations were adiabatic. 
In our view, since the theoretical situation is 
not clear cut,  such a check is probably essential in order for us to reliably 
interpret the CMB anistropy as a probe of cosmological parameters.
Our finding here is that the MAP satellite alone, even with an
optimistic assumption about the polarisation measurement, will
be unable to set useful limits on the amplitudes of isocurvature modes. 
The PLANCK satellite will be much more powerful in this respect,
and will be able to limit the amplitudes of isocurvature modes to
less than ten per cent of the adiabatic mode.

A second issue is the extent to which 
the errors
in determining cosmological parameters from the CMB alone 
degrade as an increasing number of isocurvature modes
is admitted. When in addition to the adiabatic mode
only one isocurvature mode is allowed, MAP (with polarisation) 
can still place 
quite stringent constraints on the possible contribution of the 
isocurvature modes (Tables 2-4), and the increase in the errors
in determining the other cosmological parameters is quite 
modest. However, when more 
than one isocurvature mode and the corresponding 
correlations are considered, the errors increase quite
dramatically, and worse, there are only very weak limits on the
contamination of the adiabatic mode with isocurvature modes. 
In the case of all four isocurvature modes the fractional
errors become of order one, even when polarization information
(using the optimistic estimate)
is taken into account. 
Without a polarisation measurement, admitting isocurvature modes
ruins MAP's ability to measure cosmological parameters. 
The behaviour observed is likely to be simply a result of the model
possessing too many degrees of freedom. For
all models considered the CMB moments are rather smooth, slowly varying
functions of $\ell ,$ so that a spline passing through
a rather modest number of points would quite accurately characterize any
of the theoretical models considered. Hence in regard to parameter
estimation the CMB data
in practice contains much less useful information than
one might naively conclude if one argued that 
all of the $C_\ell$'s are independent. Instead of giving us
$l_{max} \sim 1000$ numbers, we obtain more like 10.

The situation is improved if we consider the PLANCK's
estimated sensitivities. PLANCK has been designed to accurately
measure polarisation and so the estimated errors here may be 
more realistic than those for MAP. Our results demonstrate there is 
a high payoff for an accurate measurement of the polarisation.
Table 8-P-TP shows that even when we allow all possible
isocurvature modes, with arbitrary cross correlations,
then PLANCK can set upper limits of less than ten per cent 
on the isocurvature mode auto- and cross- correlation power relative to
the adiabatic power. Simultaneously, PLANCK can constrain the 
most interesting cosmological parameters to a few per cent or better. 

In conclusion this study has focussed on one difficulty in interpreting 
the CMB anisotropy data, namely in checking the assumption that
the primordial perturbations were adiabatic. We have 
made severe idealisations in other respects, namely assuming
Gaussianity and uniform noise, and in ignoring foreground contamination.
Dealing with these issues will be a massive challenge for the real
experiments. Nevertheless the calculations reported do offer a 
clear goal and a lesson, namely that high precision measurements
of the polarisation, as well as the temperature of the cosmic
microwave sky, will likely be essential to a conclusive
understanding of it.

\vskip 12pt
\noindent
{\bf Acknowledgements:} 
We would like to thank David Langlois for useful discussions. 
The CMB spectra were computed using a modified
version of the code CMBFAST written by Uros Seljak and Matias
Zaldarriaga. 
This work was supported in part by the UK Particle Physics and 
Astronomy Research Council. KM was supported by the UK Commonwealth 
Scholarship Commission.

\vfill\eject

\begin{table}
\centering

\begin{tabular}{|| c || c | c | c | c | c | c ||}
& 1 & 2 & 3 & 4 & 5 & 6 \cr
\hline
\% error &   0.15  &    0.40  &    1.30  &    2.98  &   26.38  &   48.81 \cr 
\hline
$\delta h/h$  &   17.85  &  -69.72  &  -51.48  &   -2.55  &  -41.04  &  -21.90 \cr 
$\delta \Omega _b/\Omega _b$  &  -12.13  &  -36.20  &  -31.41  &    9.60  &   64.16  &   57.86 \cr 
$\delta \Omega _k$  &  -89.17  &   -8.12  &   -8.25  &  -38.79  &   -1.46  &  -20.20 \cr 
$\delta \Omega _\Lambda /\Omega _\Lambda $  &  -39.76  &   -4.37  &    8.07  &   82.30  &  -34.96  &   18.43 \cr 
$\delta n_s/n_s$  &   -1.47  &   61.08  &  -76.84  &    1.16  &  -14.57  &   12.16 \cr 
$\tau _{reion}$  &    0.58  &   -3.56  &   18.02  &  -40.27  &  -52.56  &   72.65 \cr 
\end{tabular}
\end{table}
\begin{table}
\centering

\begin{tabular}{|| c | c | c | c | c | c ||}
$\delta h/h$ & $\delta \Omega _b/\Omega _b$ & $\delta \Omega _k$ & $\delta \Omega _\Lambda /\Omega _\Lambda $ & $\delta n_s/n_s$ & $\tau _{reion}$ \cr
\hline
  15.23  &   32.93  &    9.94  &   13.12  &    7.15  &   38.09 \cr 
\end{tabular}
\end{table}

\centerline{\bf Table 1-M-T}

\vfill\eject
\begin{table}
\centering
\begin{tabular}{|| c || c | c | c | c | c | c ||}
& 1 & 2 & 3 & 4 & 5 & 6 \cr
\hline
\% error &   0.14  &    0.40  &    1.23  &    1.76  &    3.17  &   20.22 \cr 
\hline
$\delta h/h$  &   17.64  &  -69.86  &  -47.80  &  -19.59  &    2.25  &  -46.20 \cr 
$\delta \Omega _b/\Omega _b$  &  -12.17  &  -36.19  &  -28.46  &  -11.32  &   18.48  &   85.22 \cr 
$\delta \Omega _k$  &  -89.28  &   -8.07  &   -6.81  &  -12.74  &  -40.30  &  -11.40 \cr 
$\delta \Omega _\Lambda /\Omega _\Lambda $  &  -39.58  &   -4.07  &    6.09  &   23.20  &   86.05  &  -20.92 \cr 
$\delta n_s/n_s$  &   -1.41  &   60.91  &  -69.11  &  -35.32  &   15.32  &   -5.43 \cr 
$\tau _{reion}$  &    0.71  &   -4.20  &   45.23  &  -86.83  &   19.75  &   -2.39 \cr 
\end{tabular}
\end{table}
\begin{table}
\centering

\begin{tabular}{|| c | c | c | c | c | c ||}
$\delta h/h$ & $\delta \Omega _b/\Omega _b$ & $\delta \Omega _k$ & $\delta \Omega _\Lambda /\Omega _\Lambda $ & $\delta n_s/n_s$ & $\tau _{reion}$ \cr
\hline
   9.37  &   17.25  &    2.65  &    5.05  &    1.61  &    1.81 \cr 
\end{tabular}
\end{table}
\centerline{\bf Table 1-M-TP}

\vfill\eject
\begin{table}
\centering

\begin{tabular}{|| c || c | c | c | c | c | c ||}
& 1 & 2 & 3 & 4 & 5 & 6 \cr
\hline
\% error &   0.03  &    0.23  &    0.30  &    1.81  &    5.79  &   28.28 \cr 
\hline
$\delta h/h$  &  -26.45  &  -24.44  &  -80.07  &   15.36  &   12.68  &   43.54 \cr 
$\delta \Omega _b/\Omega _b$  &    7.36  &  -11.91  &  -45.38  &   13.05  &  -16.75  &  -85.40 \cr 
$\delta \Omega _k$  &   87.22  &    4.52  &  -28.26  &  -35.55  &   -3.73  &   17.20 \cr 
$\delta \Omega _\Lambda /\Omega _\Lambda $  &   40.46  &  -20.69  &   16.16  &   85.09  &   19.64  &    6.93 \cr 
$\delta n_s/n_s$  &   -1.24  &   93.84  &  -21.68  &   25.55  &    8.37  &    0.59 \cr 
$\tau _{reion}$  &    0.00  &   -2.64  &    0.15  &  -20.91  &   95.34  &  -21.60 \cr 
\end{tabular}
\end{table}
\begin{table}
\centering

\begin{tabular}{|| c | c | c | c | c | c ||}
$\delta h/h$ & $\delta \Omega _b/\Omega _b$ & $\delta \Omega _k$ & $\delta \Omega _\Lambda /\Omega _\Lambda $ & $\delta n_s/n_s$ & $\tau _{reion}$ \cr
\hline
  12.34  &   24.18  &    4.91  &    2.74  &    0.73  &    8.24 \cr 
\end{tabular}
\end{table}
\centerline{\bf Table 1-P-T}

\vfill\eject
\begin{table}
\centering

\begin{tabular}{|| c || c | c | c | c | c | c ||}
& 1 & 2 & 3 & 4 & 5 & 6 \cr
\hline
\% error &   0.02  &    0.17  &    0.21  &    0.39  &    0.85  &    3.54 \cr 
\hline
$\delta h/h$  &   25.69  &   79.23  &   28.41  &    2.28  &   11.84  &   45.93 \cr 
$\delta \Omega _b/\Omega _b$  &   -7.20  &   42.53  &   14.76  &    0.94  &   25.82  &  -85.17 \cr 
$\delta \Omega _k$  &  -88.15  &   28.76  &   -2.41  &   -4.21  &  -35.59  &   10.56 \cr 
$\delta \Omega _\Lambda /\Omega _\Lambda $  &  -38.95  &  -20.43  &   20.50  &   10.74  &   83.89  &   22.19 \cr 
$\delta n_s/n_s$  &    0.41  &   25.76  &  -92.22  &   -3.76  &   28.08  &    5.32 \cr 
$\tau _{reion}$  &    0.03  &   -2.19  &    6.60  &  -99.23  &   10.04  &    2.00 \cr 
\end{tabular}
\end{table}
\begin{table}
\centering

\begin{tabular}{|| c | c | c | c | c | c ||}
$\delta h/h$ & $\delta \Omega _b/\Omega _b$ & $\delta \Omega _k$ & $\delta \Omega _\Lambda /\Omega _\Lambda $ & $\delta n_s/n_s$ & $\tau _{reion}$ \cr
\hline
   1.64  &    3.02  &    0.48  &    1.06  &    0.36  &    0.40 \cr 
\end{tabular}
\end{table}
\centerline{\bf Table 1-P-TP}

\vfill\eject
\begin{table}
\centering

\begin{tabular}{|| c || c | c | c | c | c | c | c | c ||}
& 1 & 2 & 3 & 4 & 5 & 6 & 7 & 8 \cr
\hline
\% error &   0.14  &    0.40  &    1.06  &    2.72  &    3.70  &    9.93  &   27.37  &   50.58 \cr 
\hline
$\delta h/h$  &   17.60  &   69.74  &  -35.52  &   18.12  &  -29.12  &  -11.82  &   43.39  &  -19.12 \cr 
$\delta \Omega _b/\Omega _b$  &  -12.17  &   35.94  &  -21.20  &    2.49  &  -26.09  &   -5.31  &  -65.74  &   55.44 \cr 
$\delta \Omega _k$  &  -88.78  &    6.76  &   -0.81  &   35.71  &    9.93  &  -17.69  &    2.88  &  -19.39 \cr 
$\delta \Omega _\Lambda /\Omega _\Lambda $  &  -39.59  &    3.67  &    7.18  &  -70.50  &  -37.94  &   20.41  &   34.13  &   19.42 \cr 
$\delta n_s/n_s$  &   -1.40  &  -59.71  &  -66.91  &   18.11  &  -33.90  &   -7.53  &   15.38  &   13.56 \cr 
$\tau _{reion}$  &    0.58  &    3.34  &   15.22  &   27.92  &   31.54  &   -2.40  &   48.69  &   74.87 \cr 
$\langle II\rangle $  &    7.22  &  -13.28  &   58.78  &   38.50  &  -69.19  &   -6.02  &    2.93  &    1.28 \cr 
$\langle AI\rangle $  &    6.17  &   -5.67  &    8.50  &  -28.81  &    1.36  &  -94.89  &    2.44  &    4.02 \cr 
\end{tabular}
\end{table}
\begin{table}
\centering

\begin{tabular}{|| c | c | c | c | c | c | c | c ||}
$\delta h/h$ & $\delta \Omega _b/\Omega _b$ & $\delta \Omega _k$ & $\delta \Omega _\Lambda /\Omega _\Lambda $ & $\delta n_s/n_s$ & $\tau _{reion}$ & $\langle II\rangle $ & $\langle AI\rangle $ \cr
\hline
  15.41  &   33.34  &   10.05  &   13.91  &    8.23  &   40.17  &    3.08  &    9.70 \cr 
\end{tabular}
\end{table}
\centerline{\bf Table 2-M-T}

\vfill\eject
\begin{table}
\centering

\begin{tabular}{|| c || c | c | c | c | c | c | c | c ||}
& 1 & 2 & 3 & 4 & 5 & 6 & 7 & 8 \cr
\hline
\% error &   0.14  &    0.40  &    1.02  &    1.72  &    2.61  &    3.56  &    8.25  &   21.00 \cr 
\hline
$\delta h/h$  &   17.40  &   69.88  &  -34.11  &    6.52  &  -26.43  &  -26.21  &   14.75  &  -44.77 \cr 
$\delta \Omega _b/\Omega _b$  &  -12.20  &   35.93  &  -20.12  &    3.64  &   -5.94  &  -28.81  &   -8.08  &   84.92 \cr 
$\delta \Omega _k$  &  -88.90  &    6.70  &   -0.41  &    8.12  &  -32.71  &   21.21  &   19.50  &   -9.29 \cr 
$\delta \Omega _\Lambda /\Omega _\Lambda $  &  -39.40  &    3.37  &    6.78  &  -20.14  &   57.57  &  -60.94  &  -20.46  &  -23.21 \cr 
$\delta n_s/n_s$  &   -1.35  &  -59.51  &  -63.75  &   20.41  &  -22.81  &  -37.54  &    4.71  &   -4.87 \cr 
$\tau _{reion}$  &    0.71  &    3.90  &   29.89  &   93.20  &   13.90  &  -14.27  &    1.53  &   -2.19 \cr 
$\langle II\rangle $  &    7.23  &  -13.38  &   57.93  &  -16.72  &  -60.28  &  -49.99  &   -1.04  &   -0.13 \cr 
$\langle AI\rangle $  &    6.20  &   -5.75  &    8.51  &   -9.47  &   23.13  &  -14.41  &   94.30  &   11.44 \cr 
\end{tabular}
\end{table}
\begin{table}
\centering

\begin{tabular}{|| c | c | c | c | c | c | c | c ||}
$\delta h/h$ & $\delta \Omega _b/\Omega _b$ & $\delta \Omega _k$ & $\delta \Omega _\Lambda /\Omega _\Lambda $ & $\delta n_s/n_s$ & $\tau _{reion}$ & $\langle II\rangle $ & $\langle AI\rangle $ \cr
\hline
   9.56  &   17.88  &    2.78  &    5.81  &    1.99  &    1.81  &    2.47  &    8.19 \cr 
\end{tabular}
\end{table}
\centerline{\bf Table 2-M-TP}

\vfill\eject
\begin{table}
\centering

\begin{tabular}{|| c || c | c | c | c | c | c | c | c ||}
& 1 & 2 & 3 & 4 & 5 & 6 & 7 & 8 \cr
\hline
\% error &   0.03  &    0.23  &    0.29  &    1.28  &    1.86  &    2.77  &    6.54  &   29.55 \cr 
\hline
$\delta h/h$  &   26.44  &   23.89  &  -76.50  &  -19.39  &  -20.05  &   -8.96  &  -10.93  &   43.59 \cr 
$\delta \Omega _b/\Omega _b$  &   -7.36  &   11.60  &  -43.18  &   -9.20  &  -14.94  &   -6.04  &   21.28  &  -84.55 \cr 
$\delta \Omega _k$  &  -87.19  &   -4.70  &  -25.98  &  -17.99  &   32.60  &   -1.64  &    3.31  &   17.36 \cr 
$\delta \Omega _\Lambda /\Omega _\Lambda $  &  -40.44  &   20.77  &   15.31  &   22.59  &  -82.49  &   -5.26  &  -17.66  &    6.60 \cr 
$\delta n_s/n_s$  &    1.24  &  -93.89  &  -19.61  &   -3.17  &  -27.24  &   -0.58  &   -6.72  &    0.54 \cr 
$\tau _{reion}$  &    0.00  &    2.65  &    0.31  &  -14.37  &   15.24  &   -2.62  &  -94.58  &  -24.54 \cr 
$\langle II\rangle $  &    2.17  &    2.76  &   29.50  &  -92.08  &  -22.51  &    5.45  &   10.06  &    1.26 \cr 
$\langle AI\rangle $  &    1.76  &   -3.27  &   10.90  &   -3.25  &    5.08  &  -99.08  &    3.67  &    1.29 \cr 
\end{tabular}
\end{table}
\begin{table}
\centering

\begin{tabular}{|| c | c | c | c | c | c | c | c ||}
$\delta h/h$ & $\delta \Omega _b/\Omega _b$ & $\delta \Omega _k$ & $\delta \Omega _\Lambda /\Omega _\Lambda $ & $\delta n_s/n_s$ & $\tau _{reion}$ & $\langle II\rangle $ & $\langle AI\rangle $ \cr
\hline
  12.91  &   25.02  &    5.17  &    2.76  &    0.73  &    9.54  &    1.47  &    2.78 \cr 
\end{tabular}
\end{table}
\centerline{\bf Table 2-P-T}

\vfill\eject
\begin{table}
\centering

\begin{tabular}{|| c || c | c | c | c | c | c | c | c ||}
& 1 & 2 & 3 & 4 & 5 & 6 & 7 & 8 \cr
\hline
\% error &   0.02  &    0.16  &    0.21  &    0.34  &    0.41  &    0.81  &    1.25  &    3.66 \cr 
\hline
$\delta h/h$  &   25.66  &   74.17  &  -30.82  &  -20.87  &   14.52  &  -13.44  &    2.38  &   45.36 \cr 
$\delta \Omega _b/\Omega _b$  &   -7.20  &   39.74  &  -16.02  &   -9.95  &    6.21  &  -25.02  &   -8.31  &  -85.32 \cr 
$\delta \Omega _k$  &  -88.10  &   25.90  &    1.36  &  -13.06  &    2.37  &   31.41  &   17.40  &   10.03 \cr 
$\delta \Omega _\Lambda /\Omega _\Lambda $  &  -38.92  &  -19.98  &  -19.72  &   10.69  &    7.26  &  -77.12  &  -32.64  &   22.75 \cr 
$\delta n_s/n_s$  &    0.41  &   25.57  &   91.10  &  -14.87  &    3.45  &  -24.71  &  -13.22  &    5.29 \cr 
$\tau _{reion}$  &    0.03  &   -3.61  &   -7.02  &  -49.24  &  -85.86  &  -11.58  &    0.29  &    2.50 \cr 
$\langle II\rangle $  &    3.01  &  -32.08  &   -5.96  &  -80.21  &   46.69  &    7.69  &  -15.81  &   -2.04 \cr 
$\langle AI\rangle $  &    2.15  &  -12.39  &    4.27  &  -10.06  &   11.33  &  -38.16  &   90.18  &   -2.36 \cr 
\end{tabular}
\end{table}
\begin{table}
\centering

\begin{tabular}{|| c | c | c | c | c | c | c | c ||}
$\delta h/h$ & $\delta \Omega _b/\Omega _b$ & $\delta \Omega _k$ & $\delta \Omega _\Lambda /\Omega _\Lambda $ & $\delta n_s/n_s$ & $\tau _{reion}$ & $\langle II\rangle $ & $\langle AI\rangle $ \cr
\hline
   1.67  &    3.13  &    0.50  &    1.12  &    0.38  &    0.41  &    0.40  &    1.17 \cr 
\end{tabular}
\end{table}
\centerline{\bf Table 2-P-TP}

\vfill\eject
\begin{table}
\centering

\begin{tabular}{|| c || c | c | c | c | c | c | c | c ||}
& 1 & 2 & 3 & 4 & 5 & 6 & 7 & 8 \cr
\hline
\% error &   0.15  &    0.40  &    0.94  &    1.80  &    8.29  &   16.18  &   49.96  &   56.13 \cr 
\hline
$\delta h/h$  &  -17.82  &  -68.13  &  -40.87  &  -31.79  &    1.82  &  -11.28  &   21.87  &   41.84 \cr 
$\delta \Omega _b/\Omega _b$  &   12.13  &  -35.23  &  -25.70  &  -14.28  &    2.56  &   48.34  &  -58.33  &  -44.74 \cr 
$\delta \Omega _k$  &   89.13  &   -7.95  &   -0.15  &  -24.34  &  -25.29  &  -15.77  &   21.21  &   -7.86 \cr 
$\delta \Omega _\Lambda /\Omega _\Lambda $  &   39.74  &   -3.87  &   -8.11  &   47.89  &   50.40  &   18.36  &  -20.36  &   52.48 \cr 
$\delta n_s/n_s$  &    1.46  &   61.11  &  -40.22  &  -58.10  &   14.39  &   13.87  &  -12.88  &   26.54 \cr 
$\tau _{reion}$  &   -0.58  &   -3.93  &   16.82  &   -7.20  &  -28.62  &  -54.35  &  -71.30  &   28.16 \cr 
$\langle II\rangle $  &   -0.89  &  -13.88  &   73.20  &  -40.45  &    2.66  &   43.80  &   -0.57  &   29.77 \cr 
$\langle AI\rangle $  &    2.86  &   -9.82  &   18.57  &  -28.91  &   76.00  &  -43.41  &   -2.39  &  -32.38 \cr 
\end{tabular}
\end{table}
\begin{table}
\centering

\begin{tabular}{|| c | c | c | c | c | c | c | c ||}
$\delta h/h$ & $\delta \Omega _b/\Omega _b$ & $\delta \Omega _k$ & $\delta \Omega _\Lambda /\Omega _\Lambda $ & $\delta n_s/n_s$ & $\tau _{reion}$ & $\langle II\rangle $ & $\langle AI\rangle $ \cr
\hline
  25.98  &   39.26  &   11.95  &   31.59  &   16.46  &   40.02  &   18.18  &   20.52 \cr 
\end{tabular}
\end{table}
\centerline{\bf Table 3-M-T}

\vfill\eject
\begin{table}
\centering

\begin{tabular}{|| c || c | c | c | c | c | c | c | c ||}
& 1 & 2 & 3 & 4 & 5 & 6 & 7 & 8 \cr
\hline
\% error &   0.14  &    0.39  &    0.91  &    1.69  &    1.91  &    7.07  &   10.07  &   24.56 \cr 
\hline
$\delta h/h$  &  -17.62  &  -68.22  &  -39.88  &  -31.83  &    8.53  &    1.24  &   -1.92  &  -48.52 \cr 
$\delta \Omega _b/\Omega _b$  &   12.17  &  -35.19  &  -24.59  &  -15.84  &   -0.89  &  -24.01  &  -36.80  &   76.33 \cr 
$\delta \Omega _k$  &   89.25  &   -7.89  &   -0.22  &  -20.90  &   13.77  &   33.47  &   14.28  &   -4.71 \cr 
$\delta \Omega _\Lambda /\Omega _\Lambda $  &   39.56  &   -3.56  &   -7.68  &   40.21  &  -29.47  &  -62.54  &  -27.05  &  -35.13 \cr 
$\delta n_s/n_s$  &    1.40  &   60.94  &  -37.49  &  -57.91  &   14.83  &  -24.36  &  -22.39  &  -14.51 \cr 
$\tau _{reion}$  &   -0.71  &   -4.81  &   30.81  &  -47.84  &  -81.96  &    4.18  &    1.02  &   -1.26 \cr 
$\langle II\rangle $  &   -0.82  &  -14.05  &   71.08  &  -23.02  &   39.94  &  -13.38  &  -47.21  &  -14.73 \cr 
$\langle AI\rangle $  &    2.88  &   -9.97  &   18.42  &  -22.58  &   18.30  &  -60.00  &   70.54  &   11.53 \cr 
\end{tabular}
\end{table}
\begin{table}
\centering

\begin{tabular}{|| c | c | c | c | c | c | c | c ||}
$\delta h/h$ & $\delta \Omega _b/\Omega _b$ & $\delta \Omega _k$ & $\delta \Omega _\Lambda /\Omega _\Lambda $ & $\delta n_s/n_s$ & $\tau _{reion}$ & $\langle II\rangle $ & $\langle AI\rangle $ \cr
\hline
  11.94  &   19.19  &    3.04  &   10.11  &    4.69  &    1.84  &    6.14  &    8.76 \cr 
\end{tabular}
\end{table}
\centerline{\bf Table 3-M-TP}

\vfill\eject
\begin{table}
\centering

\begin{tabular}{|| c || c | c | c | c | c | c | c | c ||}
& 1 & 2 & 3 & 4 & 5 & 6 & 7 & 8 \cr
\hline
\% error &   0.03  &    0.23  &    0.30  &    1.07  &    2.20  &    4.70  &   15.88  &   36.36 \cr 
\hline
$\delta h/h$  &  -26.45  &  -24.52  &   79.75  &    7.96  &   14.23  &    7.83  &   21.64  &  -39.29 \cr 
$\delta \Omega _b/\Omega _b$  &    7.36  &  -11.94  &   45.21  &    8.80  &    6.53  &    5.67  &  -44.41  &   75.07 \cr 
$\delta \Omega _k$  &   87.22  &    4.35  &   28.21  &  -14.14  &  -32.60  &   -4.24  &    8.22  &  -15.17 \cr 
$\delta \Omega _\Lambda /\Omega _\Lambda $  &   40.46  &  -20.25  &  -16.28  &   34.56  &   78.86  &   14.46  &    4.41  &   -6.77 \cr 
$\delta n_s/n_s$  &   -1.24  &   92.82  &   22.31  &   -2.39  &   29.49  &    2.93  &    0.91  &   -0.52 \cr 
$\tau _{reion}$  &    0.00  &   -2.73  &   -0.17  &  -19.68  &    3.10  &   23.78  &   81.83  &   48.31 \cr 
$\langle II\rangle $  &    0.09  &  -11.93  &   -1.54  &  -86.07  &   29.39  &   29.44  &  -24.70  &  -10.28 \cr 
$\langle AI\rangle $  &    0.55  &   -7.88  &    5.97  &  -25.75  &   27.02  &  -90.77  &   12.87  &   10.23 \cr 
\end{tabular}
\end{table}
\begin{table}
\centering

\begin{tabular}{|| c | c | c | c | c | c | c | c ||}
$\delta h/h$ & $\delta \Omega _b/\Omega _b$ & $\delta \Omega _k$ & $\delta \Omega _\Lambda /\Omega _\Lambda $ & $\delta n_s/n_s$ & $\tau _{reion}$ & $\langle II\rangle $ & $\langle AI\rangle $ \cr
\hline
  14.70  &   28.20  &    5.72  &    3.19  &    0.74  &   21.88  &    5.71  &    6.05 \cr 
\end{tabular}
\end{table}
\centerline{\bf Table 3-P-T}

\vfill\eject
\begin{table}
\centering

\begin{tabular}{|| c || c | c | c | c | c | c | c | c ||}
& 1 & 2 & 3 & 4 & 5 & 6 & 7 & 8 \cr
\hline
\% error &   0.02  &    0.17  &    0.21  &    0.38  &    0.65  &    0.98  &    1.58  &    3.80 \cr 
\hline
$\delta h/h$  &  -25.68  &  -79.31  &   26.65  &    4.34  &  -11.77  &   -8.63  &  -11.10  &  -44.56 \cr 
$\delta \Omega _b/\Omega _b$  &    7.20  &  -42.56  &   13.74  &    2.43  &  -15.08  &  -20.48  &   18.62  &   83.36 \cr 
$\delta \Omega _k$  &   88.15  &  -28.65  &   -2.55  &   -5.14  &   18.08  &   27.66  &  -14.63  &   -8.34 \cr 
$\delta \Omega _\Lambda /\Omega _\Lambda $  &   38.95  &   20.27  &   19.65  &   14.04  &  -46.59  &  -64.98  &   20.85  &  -25.72 \cr 
$\delta n_s/n_s$  &   -0.41  &  -24.84  &  -90.82  &   -8.41  &   -1.55  &  -29.35  &   11.91  &   -7.58 \cr 
$\tau _{reion}$  &   -0.03  &    2.12  &    7.78  &  -95.41  &  -28.52  &    3.88  &    1.00  &   -1.85 \cr 
$\langle II\rangle $  &    0.02  &    1.82  &   16.35  &  -22.82  &   71.93  &  -59.25  &  -22.77  &    2.44 \cr 
$\langle AI\rangle $  &    0.71  &   -6.55  &   11.46  &   -7.60  &   33.87  &   11.48  &   90.67  &  -16.38 \cr 
\end{tabular}
\end{table}
\begin{table}
\centering

\begin{tabular}{|| c | c | c | c | c | c | c | c ||}
$\delta h/h$ & $\delta \Omega _b/\Omega _b$ & $\delta \Omega _k$ & $\delta \Omega _\Lambda /\Omega _\Lambda $ & $\delta n_s/n_s$ & $\tau _{reion}$ & $\langle II\rangle $ & $\langle AI\rangle $ \cr
\hline
   1.71  &    3.19  &    0.49  &    1.25  &    0.49  &    0.41  &    0.84  &    1.58 \cr 
\end{tabular}
\end{table}
\centerline{\bf Table 3-P-TP}

\vfill\eject
\begin{table}
\centering

\begin{tabular}{|| c || c | c | c | c | c | c | c | c ||}
& 1 & 2 & 3 & 4 & 5 & 6 & 7 & 8 \cr
\hline
\% error &   0.15  &    0.40  &    1.00  &    1.76  &    6.28  &    7.68  &   39.52  &   58.91 \cr 
\hline
$\delta h/h$  &  -17.90  &   69.48  &  -27.55  &  -40.79  &    2.87  &    5.80  &  -25.11  &   41.91 \cr 
$\delta \Omega _b/\Omega _b$  &   12.06  &   36.29  &  -19.39  &  -19.54  &   -4.06  &   33.22  &   20.01  &  -79.11 \cr 
$\delta \Omega _k$  &   89.00  &    8.64  &   -1.92  &  -22.19  &  -10.59  &  -31.96  &   13.35  &   13.98 \cr 
$\delta \Omega _\Lambda /\Omega _\Lambda $  &   39.67  &    4.86  &   -8.73  &   46.61  &   17.99  &   58.83  &  -47.48  &   10.68 \cr 
$\delta n_s/n_s$  &    1.50  &  -60.14  &  -56.64  &  -48.59  &   -4.20  &   17.37  &  -22.16  &    0.43 \cr 
$\tau _{reion}$  &   -0.56  &    3.27  &   18.46  &   -6.96  &   -2.61  &  -45.67  &  -76.92  &  -39.89 \cr 
$\langle II\rangle $  &    3.81  &   -9.74  &   69.87  &  -46.76  &  -26.05  &   44.60  &   -9.31  &    8.25 \cr 
$\langle AI\rangle $  &    4.61  &   -5.81  &   18.81  &  -26.34  &   94.00  &   -1.74  &    6.51  &   -3.97 \cr 
\end{tabular}
\end{table}
\begin{table}
\centering

\begin{tabular}{|| c | c | c | c | c | c | c | c ||}
$\delta h/h$ & $\delta \Omega _b/\Omega _b$ & $\delta \Omega _k$ & $\delta \Omega _\Lambda /\Omega _\Lambda $ & $\delta n_s/n_s$ & $\tau _{reion}$ & $\langle II\rangle $ & $\langle AI\rangle $ \cr
\hline
  26.63  &   47.35  &   10.12  &   20.35  &    8.93  &   38.58  &    7.26  &    6.87 \cr 
\end{tabular}
\end{table}
\centerline{\bf Table 4-M-T}
 
\vfill\eject
\begin{table}
\centering

\begin{tabular}{|| c || c | c | c | c | c | c | c | c ||}
& 1 & 2 & 3 & 4 & 5 & 6 & 7 & 8 \cr
\hline
\% error &   0.14  &    0.40  &    0.97  &    1.67  &    1.84  &    5.54  &    6.40  &   29.31 \cr 
\hline
$\delta h/h$  &  -17.69  &   69.64  &  -26.23  &  -40.98  &    7.05  &    0.75  &    5.14  &   48.92 \cr 
$\delta \Omega _b/\Omega _b$  &   12.10  &   36.29  &  -18.13  &  -20.06  &   -0.27  &    1.04  &   41.24  &  -78.12 \cr 
$\delta \Omega _k$  &   89.12  &    8.58  &   -1.66  &  -21.47  &    5.62  &    9.02  &  -37.29  &    4.10 \cr 
$\delta \Omega _\Lambda /\Omega _\Lambda $  &   39.49  &    4.57  &   -8.94  &   44.18  &  -15.85  &  -12.40  &   68.87  &   35.22 \cr 
$\delta n_s/n_s$  &    1.43  &  -59.99  &  -52.99  &  -52.93  &   -5.16  &    6.30  &   24.43  &   11.25 \cr 
$\tau _{reion}$  &   -0.69  &    3.80  &   33.16  &  -27.55  &  -90.08  &    2.88  &    0.95  &    1.89 \cr 
$\langle II\rangle $  &    3.93  &   -9.66  &   68.07  &  -36.92  &   37.48  &   28.92  &   39.23  &   10.77 \cr 
$\langle AI\rangle $  &    4.61  &   -5.68  &   18.97  &  -24.12  &   11.08  &  -94.23  &    1.56  &   -0.60 \cr 
\end{tabular}
\end{table}
\begin{table}
\centering

\begin{tabular}{|| c | c | c | c | c | c | c | c ||}
$\delta h/h$ & $\delta \Omega _b/\Omega _b$ & $\delta \Omega _k$ & $\delta \Omega _\Lambda /\Omega _\Lambda $ & $\delta n_s/n_s$ & $\tau _{reion}$ & $\langle II\rangle $ & $\langle AI\rangle $ \cr
\hline
  14.37  &   23.06  &    2.75  &   11.28  &    3.81  &    1.84  &    4.49  &    5.25 \cr 
\end{tabular}
\end{table}
\centerline{\bf Table 4-M-TP}
 
\vfill\eject
\begin{table}
\centering

\begin{tabular}{|| c || c | c | c | c | c | c | c | c ||}
& 1 & 2 & 3 & 4 & 5 & 6 & 7 & 8 \cr
\hline
\% error &   0.03  &    0.23  &    0.30  &    0.92  &    1.93  &    2.96  &    6.55  &   30.58 \cr 
\hline
$\delta h/h$  &  -26.44  &  -24.37  &  -78.56  &  -12.35  &   18.58  &    3.04  &  -11.67  &  -43.49 \cr 
$\delta \Omega _b/\Omega _b$  &    7.35  &  -11.87  &  -44.68  &   -4.50  &   12.84  &    6.49  &   20.29  &   84.67 \cr 
$\delta \Omega _k$  &   87.18  &    4.57  &  -28.10  &  -10.81  &  -33.33  &   -8.57  &    1.34  &  -16.94 \cr 
$\delta \Omega _\Lambda /\Omega _\Lambda $  &   40.44  &  -20.69  &   15.33  &   22.72  &   79.94  &   22.68  &  -15.04  &   -7.37 \cr 
$\delta n_s/n_s$  &   -1.24  &   93.85  &  -21.22  &    1.12  &   25.78  &    5.67  &   -6.49  &   -0.67 \cr 
$\tau _{reion}$  &    0.00  &   -2.64  &    0.35  &  -11.33  &  -12.08  &   -4.15  &  -95.47  &   24.24 \cr 
$\langle II\rangle $  &    2.42  &   -0.03  &   18.27  &  -92.87  &   15.18  &   27.29  &    7.55  &   -1.71 \cr 
$\langle AI\rangle $  &    2.11  &   -1.27  &    4.72  &  -20.96  &   30.78  &  -92.56  &    3.34  &    2.64 \cr 
\end{tabular}
\end{table}
\begin{table}
\centering

\begin{tabular}{|| c | c | c | c | c | c | c | c ||}
$\delta h/h$ & $\delta \Omega _b/\Omega _b$ & $\delta \Omega _k$ & $\delta \Omega _\Lambda /\Omega _\Lambda $ & $\delta n_s/n_s$ & $\tau _{reion}$ & $\langle II\rangle $ & $\langle AI\rangle $ \cr
\hline
  13.33  &   25.93  &    5.23  &    2.99  &    0.74  &    9.70  &    1.41  &    2.93 \cr 
\end{tabular}
\end{table}
\centerline{\bf Table 4-P-T}

\vfill\eject
\begin{table}
\centering

\begin{tabular}{|| c || c | c | c | c | c | c | c | c ||}
& 1 & 2 & 3 & 4 & 5 & 6 & 7 & 8 \cr
\hline
\% error &   0.02  &    0.17  &    0.21  &    0.30  &    0.39  &    0.89  &    1.11  &    3.72 \cr 
\hline
$\delta h/h$  &  -25.67  &  -78.84  &   29.26  &    3.95  &    1.64  &   12.67  &    0.37  &   45.72 \cr 
$\delta \Omega _b/\Omega _b$  &    7.20  &  -42.40  &   14.77  &    5.85  &   -0.45  &   19.31  &   19.51  &  -84.52 \cr 
$\delta \Omega _k$  &   88.10  &  -28.78  &   -1.86  &   -5.32  &   -2.71  &  -29.14  &  -20.63  &    9.84 \cr 
$\delta \Omega _\Lambda /\Omega _\Lambda $  &   38.92  &   20.38  &   17.97  &   21.07  &    5.98  &   72.01  &   39.29  &   23.18 \cr 
$\delta n_s/n_s$  &   -0.41  &  -26.72  &  -89.92  &  -16.88  &   -2.55  &   24.16  &   17.04  &    5.95 \cr 
$\tau _{reion}$  &   -0.03  &    2.35  &    7.16  &  -12.93  &  -98.54  &    7.42  &    2.69  &    2.02 \cr 
$\langle II\rangle $  &    2.69  &    4.88  &   19.82  &  -92.45  &   15.21  &   25.38  &  -12.19  &   -2.25 \cr 
$\langle AI\rangle $  &    2.33  &   -0.34  &    8.37  &  -21.89  &    2.49  &  -46.35  &   84.88  &    9.31 \cr 
\end{tabular}
\end{table}
\begin{table}
\centering

\begin{tabular}{|| c | c | c | c | c | c | c | c ||}
$\delta h/h$ & $\delta \Omega _b/\Omega _b$ & $\delta \Omega _k$ & $\delta \Omega _\Lambda /\Omega _\Lambda $ & $\delta n_s/n_s$ & $\tau _{reion}$ & $\langle II\rangle $ & $\langle AI\rangle $ \cr
\hline
   1.71  &    3.16  &    0.51  &    1.16  &    0.41  &    0.40  &    0.40  &    1.09 \cr 
\end{tabular}
\end{table}
\centerline{\bf Table 4-P-TP}

\vfill\eject
\begin{table}
\centering

\begin{tabular}{|| c || c | c | c | c | c | c | c | c | c | c | c ||}
& 1 & 2 & 3 & 4 & 5 & 6 & 7 & 8 & 9 & 10 & 11 \cr
\hline
\% error &   0.15  &    0.39  &    0.67  &    1.64  &    4.75  &    6.53  &   16.27  &   29.15  &   41.78  &  128.96  &  142.34 \cr 
\hline
$\delta h/h$  &  -17.86  &  -66.89  &  -27.94  &  -43.27  &    6.66  &   -4.95  &   -3.99  &    3.57  &   28.55  &  -39.20  &   10.24 \cr 
$\delta \Omega _b/\Omega _b$  &   12.07  &  -34.63  &  -18.23  &  -22.19  &   14.46  &  -19.72  &   21.74  &  -19.11  &  -48.63  &   61.11  &   17.17 \cr 
$\delta \Omega _k$  &   88.97  &   -8.20  &   -1.50  &  -20.02  &    3.67  &   28.27  &  -14.75  &    3.95  &   -3.35  &  -12.19  &  -20.22 \cr 
$\delta \Omega _\Lambda /\Omega _\Lambda $  &   39.65  &   -3.88  &  -10.47  &   40.09  &   -6.91  &  -47.61  &   29.77  &   -4.27  &   35.72  &   -8.76  &   46.09 \cr 
$\delta n_s/n_s$  &    1.47  &   60.62  &  -21.85  &  -65.50  &    0.49  &   -2.25  &   24.63  &   -9.14  &   19.57  &    0.74  &   21.82 \cr 
$\tau _{reion}$  &   -0.56  &   -4.02  &   12.03  &   -0.24  &   -2.70  &   28.40  &  -58.84  &  -31.87  &   32.02  &   33.89  &   48.71 \cr 
$\langle I_1I_1\rangle $  &   -0.82  &  -14.32  &   53.52  &  -10.44  &  -30.99  &   36.78  &   40.85  &   -2.22  &  -23.99  &  -22.64  &   41.78 \cr 
$\langle I_2I_2\rangle $  &    3.82  &    6.29  &   50.34  &  -18.60  &   59.47  &  -35.46  &  -22.34  &   33.72  &  -10.28  &   -9.97  &   20.77 \cr 
$\langle I_1A\rangle $  &    2.87  &   -9.87  &   14.34  &  -19.13  &  -50.19  &  -13.78  &   -4.43  &   64.83  &   24.31  &   41.82  &   -7.18 \cr 
$\langle I_2A\rangle $  &    4.61  &    4.65  &   15.71  &  -19.24  &  -50.07  &  -52.92  &  -36.20  &  -36.66  &  -27.26  &  -23.61  &   -8.81 \cr 
$\langle I_1I_2\rangle $  &    0.95  &  -13.79  &   47.99  &  -12.08  &   12.45  &  -10.40  &   28.99  &  -42.43  &   45.92  &   20.87  &  -43.73 \cr 
\end{tabular}
\end{table}
\begin{table}
\centering

\begin{tabular}{|| c | c | c | c | c | c | c | c | c | c | c ||}
$\delta h/h$ & $\delta \Omega _b/\Omega _b$ & $\delta \Omega _k$ & $\delta \Omega _\Lambda /\Omega _\Lambda $ & $\delta n_s/n_s$ & $\tau _{reion}$ & $\langle I_1I_1\rangle $ & $\langle I_2I_2\rangle $ & $\langle I_1A\rangle $ & $\langle I_2A\rangle $ & $\langle I_1I_2\rangle $ \cr
\hline
  53.97  &   85.25  &   32.98  &   68.48  &   32.52  &   84.13  &   67.40  &   34.36  &   58.99  &   37.16  &   71.71 \cr 
\end{tabular}
\end{table}
\centerline{\bf Table 5-M-T}

\vfill\eject
\begin{table}
\centering

\begin{tabular}{|| c || c | c | c | c | c | c | c | c | c | c | c ||}
& 1 & 2 & 3 & 4 & 5 & 6 & 7 & 8 & 9 & 10 & 11 \cr
\hline
\% error &   0.14  &    0.39  &    0.65  &    1.58  &    1.80  &    4.42  &    5.75  &    8.54  &   26.58  &   29.25  &   52.80 \cr 
\hline
$\delta h/h$  &  -17.65  &  -66.93  &  -27.92  &  -43.39  &   -2.31  &   -7.79  &    2.86  &   -2.26  &   33.79  &  -23.10  &   28.16 \cr 
$\delta \Omega _b/\Omega _b$  &   12.11  &  -34.56  &  -17.96  &  -21.61  &   -3.91  &  -11.49  &   29.66  &  -23.91  &  -65.37  &   36.63  &  -25.60 \cr 
$\delta \Omega _k$  &   89.08  &   -8.11  &   -1.64  &  -20.12  &   -0.78  &   -4.48  &  -31.48  &   23.01  &    2.78  &   -6.46  &   -0.95 \cr 
$\delta \Omega _\Lambda /\Omega _\Lambda $  &   39.46  &   -3.55  &  -10.38  &   40.88  &   -0.87  &    5.29  &   52.17  &  -51.21  &   26.92  &   -7.93  &   22.17 \cr 
$\delta n_s/n_s$  &    1.40  &   60.48  &  -20.41  &  -64.74  &  -11.01  &   -4.33  &    5.85  &  -37.91  &    8.29  &   -6.00  &    4.04 \cr 
$\tau _{reion}$  &   -0.68  &   -4.98  &   18.72  &    4.59  &  -97.60  &   -6.98  &    2.45  &    4.26  &    1.17  &   -1.92  &    0.56 \cr 
$\langle I_1I_1\rangle $  &   -0.75  &  -14.58  &   52.90  &   -8.65  &    5.52  &   22.84  &  -39.91  &  -48.75  &  -19.52  &   11.11  &   43.45 \cr 
$\langle I_2I_2\rangle $  &    3.94  &    5.98  &   50.11  &  -20.31  &   13.97  &  -51.68  &   40.98  &   26.80  &   12.31  &   28.34  &   28.40 \cr 
$\langle I_1A\rangle $  &    2.89  &  -10.03  &   14.40  &  -18.62  &   -1.75  &   52.19  &    9.74  &    2.08  &   47.24  &   56.11  &  -33.63 \cr 
$\langle I_2A\rangle $  &    4.61  &    4.45  &   15.72  &  -19.72  &    0.57  &   59.92  &   43.82  &   33.48  &  -30.68  &  -36.58  &   20.32 \cr 
$\langle I_1I_2\rangle $  &    1.04  &  -14.02  &   47.50  &  -11.82  &   10.13  &  -13.59  &    8.52  &  -24.42  &    8.80  &  -50.67  &  -61.68 \cr 
\end{tabular}
\end{table}
\begin{table}
\centering

\begin{tabular}{|| c | c | c | c | c | c | c | c | c | c | c ||}
$\delta h/h$ & $\delta \Omega _b/\Omega _b$ & $\delta \Omega _k$ & $\delta \Omega _\Lambda /\Omega _\Lambda $ & $\delta n_s/n_s$ & $\tau _{reion}$ & $\langle I_1I_1\rangle $ & $\langle I_2I_2\rangle $ & $\langle I_1A\rangle $ & $\langle I_2A\rangle $ & $\langle I_1I_2\rangle $ \cr
\hline
  18.66  &   24.64  &    3.42  &   14.91  &    4.93  &    1.96  &   24.24  &   17.91  &   27.35  &   17.82  &   35.93 \cr 
\end{tabular}
\end{table}
\centerline{\bf Table 5-M-TP}

\vfill\eject
\begin{table}
\centering

\begin{tabular}{|| c || c | c | c | c | c | c | c | c | c | c | c ||}
& 1 & 2 & 3 & 4 & 5 & 6 & 7 & 8 & 9 & 10 & 11 \cr
\hline
\% error &   0.03  &    0.23  &    0.30  &    0.67  &    1.70  &    2.13  &    2.95  &    9.35  &   12.27  &   20.47  &   57.63 \cr 
\hline
$\delta h/h$  &  -26.44  &  -23.83  &  -78.51  &   -5.81  &  -17.40  &   -6.33  &    8.72  &    6.24  &  -10.14  &  -16.19  &  -41.39 \cr 
$\delta \Omega _b/\Omega _b$  &    7.35  &  -11.52  &  -44.65  &   -0.96  &  -13.62  &   -0.09  &    5.85  &  -25.70  &    3.35  &   36.95  &   74.58 \cr 
$\delta \Omega _k$  &   87.17  &    4.50  &  -27.97  &   -9.38  &   18.95  &   24.38  &  -16.10  &    3.90  &   -2.05  &   -6.40  &  -15.41 \cr 
$\delta \Omega _\Lambda /\Omega _\Lambda $  &   40.44  &  -19.95  &   15.13  &   21.57  &  -45.51  &  -57.17  &   43.18  &    1.22  &   -1.29  &   -3.77  &   -7.15 \cr 
$\delta n_s/n_s$  &   -1.24  &   92.03  &  -20.72  &  -13.06  &   -9.67  &  -24.70  &   14.46  &    0.58  &    3.29  &   -2.39  &    0.06 \cr 
$\tau _{reion}$  &    0.00  &   -2.81  &    0.38  &  -10.41  &   15.53  &   -5.60  &   11.71  &   12.90  &  -58.43  &  -65.00  &   40.84 \cr 
$\langle I_1I_1\rangle $  &    0.09  &  -12.31  &    2.58  &  -48.40  &   49.39  &  -37.29  &   21.20  &   12.15  &  -27.29  &   46.86  &  -11.48 \cr 
$\langle I_2I_2\rangle $  &    2.42  &   -1.08  &   18.39  &  -62.18  &  -59.02  &   40.59  &   12.86  &    4.97  &  -18.70  &    9.92  &   -4.31 \cr 
$\langle I_1A\rangle $  &    0.55  &   -7.94  &   -5.41  &  -18.77  &  -11.25  &  -21.81  &  -31.33  &   77.80  &   36.47  &   -5.66  &   23.59 \cr 
$\langle I_2A\rangle $  &    2.11  &   -1.54  &    4.72  &  -15.34  &  -21.53  &  -41.48  &  -75.19  &  -34.59  &  -25.70  &   -0.40  &   -6.47 \cr 
$\langle I_1I_2\rangle $  &    0.41  &  -13.65  &    2.39  &  -48.33  &   15.33  &  -14.86  &   13.06  &  -41.18  &   58.05  &  -41.96  &    4.41 \cr 
\end{tabular}
\end{table}
\begin{table}
\centering

\begin{tabular}{|| c | c | c | c | c | c | c | c | c | c | c ||}
$\delta h/h$ & $\delta \Omega _b/\Omega _b$ & $\delta \Omega _k$ & $\delta \Omega _\Lambda /\Omega _\Lambda $ & $\delta n_s/n_s$ & $\tau _{reion}$ & $\langle I_1I_1\rangle $ & $\langle I_2I_2\rangle $ & $\langle I_1A\rangle $ & $\langle I_2A\rangle $ & $\langle I_1I_2\rangle $ \cr
\hline
  24.13  &   43.71  &    9.02  &    4.62  &    0.97  &   28.01  &   12.25  &    4.22  &   16.13  &    6.34  &   12.09 \cr 
\end{tabular}
\end{table}
\centerline{\bf Table 5-P-T}

\vfill\eject
\begin{table}
\centering

\begin{tabular}{|| c || c | c | c | c | c | c | c | c | c | c | c ||}
& 1 & 2 & 3 & 4 & 5 & 6 & 7 & 8 & 9 & 10 & 11 \cr
\hline
\% error &   0.02  &    0.17  &    0.19  &    0.27  &    0.39  &    0.80  &    1.00  &    1.18  &    3.47  &    4.84  &    5.62 \cr 
\hline
$\delta h/h$  &  -25.67  &   78.65  &   25.86  &   12.84  &    1.72  &  -11.00  &   -8.96  &    8.86  &  -37.95  &   18.08  &   16.50 \cr 
$\delta \Omega _b/\Omega _b$  &    7.20  &   42.30  &   12.51  &    9.46  &   -0.54  &  -11.73  &  -21.12  &  -15.57  &   64.05  &  -42.47  &  -34.35 \cr 
$\delta \Omega _k$  &   88.10  &   28.67  &   -0.93  &   -5.08  &   -2.61  &   16.07  &   29.60  &   12.41  &   -8.54  &    2.17  &    4.16 \cr 
$\delta \Omega _\Lambda /\Omega _\Lambda $  &   38.92  &  -20.34  &   12.47  &   24.03  &    5.58  &  -45.42  &  -65.17  &  -20.01  &  -16.47  &   16.05  &    8.26 \cr 
$\delta n_s/n_s$  &   -0.41  &   26.68  &  -77.89  &  -45.41  &   -3.09  &  -16.56  &  -19.65  &  -20.21  &   -2.86  &    7.24  &    4.54 \cr 
$\tau _{reion}$  &   -0.03  &   -2.42  &    9.07  &   -9.57  &  -97.64  &   -0.71  &  -12.18  &   11.19  &    0.40  &   -0.73  &    3.53 \cr 
$\langle I_1I_1\rangle $  &    0.03  &   -2.82  &   22.43  &  -22.67  &   -8.12  &  -26.97  &   34.18  &  -56.12  &  -34.85  &   -1.06  &  -51.47 \cr 
$\langle I_2I_2\rangle $  &    2.69  &   -5.16  &   33.05  &  -70.27  &   18.15  &   -8.49  &  -25.85  &   48.80  &   -1.87  &    1.38  &  -21.92 \cr 
$\langle I_1A\rangle $  &    0.71  &    6.02  &   13.97  &   -8.44  &   -0.81  &   40.67  &  -13.29  &  -27.95  &   33.42  &   75.92  &  -14.26 \cr 
$\langle I_2A\rangle $  &    2.33  &    0.40  &   10.95  &  -14.73  &    3.37  &   65.04  &  -37.74  &  -33.91  &  -31.30  &  -41.83  &   10.58 \cr 
$\langle I_1I_2\rangle $  &    0.43  &   -1.21  &   30.98  &  -35.35  &    2.97  &  -21.50  &   19.45  &  -33.61  &   28.09  &   -6.06  &   70.60 \cr 
\end{tabular}
\end{table}
\begin{table}
\centering

\begin{tabular}{|| c | c | c | c | c | c | c | c | c | c | c ||}
$\delta h/h$ & $\delta \Omega _b/\Omega _b$ & $\delta \Omega _k$ & $\delta \Omega _\Lambda /\Omega _\Lambda $ & $\delta n_s/n_s$ & $\tau _{reion}$ & $\langle I_1I_1\rangle $ & $\langle I_2I_2\rangle $ & $\langle I_1A\rangle $ & $\langle I_2A\rangle $ & $\langle I_1I_2\rangle $ \cr
\hline
   1.85  &    3.60  &    0.53  &    1.33  &    0.59  &    0.47  &    3.23  &    1.40  &    3.96  &    2.49  &    4.12 \cr 
\end{tabular}
\end{table}
\centerline{\bf Table 5-P-TP}

\vfill\eject
\begin{table}
\centering

\begin{tabular}{|| c || c | c | c | c | c | c | c | c | c | c | c ||}
& 1 & 2 & 3 & 4 & 5 & 6 & 7 & 8 & 9 & 10 & 11 \cr
\hline
\% error &   0.14  &    0.39  &    0.82  &    1.65  &    2.93  &    7.31  &    8.18  &   16.75  &   58.28  &   67.13  &  389.21 \cr 
\hline
$\delta h/h$  &  -17.57  &  -68.38  &   32.07  &  -15.03  &  -33.67  &   -1.01  &   17.24  &    0.21  &  -21.97  &  -30.74  &  -30.02 \cr 
$\delta \Omega _b/\Omega _b$  &   12.17  &  -35.10  &   19.42  &   -3.63  &  -22.58  &    0.73  &  -13.10  &  -35.81  &    1.86  &   46.21  &   64.23 \cr 
$\delta \Omega _k$  &   88.74  &   -6.64  &   -2.79  &  -17.80  &  -14.83  &   24.12  &   17.04  &   11.18  &   15.83  &    0.23  &  -17.01 \cr 
$\delta \Omega _\Lambda /\Omega _\Lambda $  &   39.56  &   -3.24  &    2.84  &   45.56  &   15.82  &  -45.38  &  -15.02  &  -21.84  &  -50.34  &  -27.78  &    4.74 \cr 
$\delta n_s/n_s$  &    1.39  &   59.82  &   41.96  &  -44.33  &  -35.05  &   -8.63  &    1.58  &  -19.39  &  -30.22  &   -9.69  &    2.36 \cr 
$\tau _{reion}$  &   -0.58  &   -3.69  &  -14.25  &  -10.57  &    5.39  &   24.15  &    8.65  &   59.71  &  -65.86  &   16.96  &   28.18 \cr 
$\langle I_1I_1\rangle $  &   -7.23  &   12.30  &  -44.72  &   28.01  &  -61.74  &  -21.73  &   50.90  &   -0.71  &    2.56  &    0.29  &   11.45 \cr 
$\langle I_2I_2\rangle $  &   -0.92  &  -12.86  &  -61.79  &  -53.61  &   11.78  &    3.13  &   -9.39  &  -36.74  &  -11.91  &  -34.26  &   15.50 \cr 
$\langle I_1A\rangle $  &   -6.17  &    5.62  &   -3.54  &   19.46  &   17.24  &   49.00  &   33.50  &  -53.08  &  -32.68  &   29.44  &  -31.08 \cr 
$\langle I_2A\rangle $  &    2.86  &   -9.55  &  -14.65  &  -31.77  &    9.30  &  -57.68  &    3.15  &    2.56  &  -10.51  &   59.65  &  -39.73 \cr 
$\langle I_1I_2\rangle $  &    1.46  &   -2.57  &   23.88  &  -15.65  &   47.90  &  -22.14  &   71.60  &   -0.23  &   12.29  &  -12.96  &   30.74 \cr 
\end{tabular}
\end{table}
\begin{table}
\centering

\begin{tabular}{|| c | c | c | c | c | c | c | c | c | c | c ||}
$\delta h/h$ & $\delta \Omega _b/\Omega _b$ & $\delta \Omega _k$ & $\delta \Omega _\Lambda /\Omega _\Lambda $ & $\delta n_s/n_s$ & $\tau _{reion}$ & $\langle I_1I_1\rangle $ & $\langle I_2I_2\rangle $ & $\langle I_1A\rangle $ & $\langle I_2A\rangle $ & $\langle I_1I_2\rangle $ \cr
\hline
 119.35  &  252.00  &   66.92  &   39.69  &   21.20  &  117.20  &   44.85  &   65.23  &  124.46  &  159.92  &  120.35 \cr 
\end{tabular}
\end{table}
\centerline{\bf Table 6-M-T}

\vfill\eject
\begin{table}
\centering

\begin{tabular}{|| c || c | c | c | c | c | c | c | c | c | c | c ||}
& 1 & 2 & 3 & 4 & 5 & 6 & 7 & 8 & 9 & 10 & 11 \cr
\hline
\% error &   0.14  &    0.39  &    0.80  &    1.52  &    1.88  &    2.61  &    5.75  &    6.61  &    9.66  &   26.39  &   37.23 \cr 
\hline
$\delta h/h$  &  -17.37  &  -68.47  &   31.55  &  -11.53  &  -10.42  &   35.39  &   13.69  &    3.46  &    7.17  &  -47.45  &    4.31 \cr 
$\delta \Omega _b/\Omega _b$  &   12.21  &  -35.06  &   18.93  &   -3.57  &   -0.39  &   20.74  &   -5.67  &  -18.60  &   37.35  &   77.76  &   -0.44 \cr 
$\delta \Omega _k$  &   88.85  &   -6.56  &   -2.78  &  -13.31  &  -13.52  &   15.67  &    8.73  &   32.63  &  -16.91  &   -4.52  &    1.51 \cr 
$\delta \Omega _\Lambda /\Omega _\Lambda $  &   39.38  &   -2.92  &    2.71  &   37.18  &   29.88  &  -15.70  &   -2.89  &  -56.37  &   39.07  &  -34.56  &    2.00 \cr 
$\delta n_s/n_s$  &    1.34  &   59.62  &   40.36  &  -41.21  &  -17.45  &   36.17  &    4.59  &  -16.85  &   31.22  &  -13.44  &    6.56 \cr 
$\tau _{reion}$  &   -0.71  &   -4.48  &  -23.73  &  -49.18  &   80.51  &   20.43  &   -8.59  &    3.03  &   -2.70  &   -2.03  &    1.91 \cr 
$\langle I_1I_1\rangle $  &   -7.24  &   12.38  &  -44.42  &   32.68  &   -1.32  &   60.00  &   53.69  &  -12.79  &   -3.78  &    5.65  &   -7.63 \cr 
$\langle I_2I_2\rangle $  &   -0.85  &  -13.00  &  -61.09  &  -39.67  &  -37.32  &  -16.78  &   -4.84  &   -4.04  &   47.39  &  -12.59  &  -20.15 \cr 
$\langle I_1A\rangle $  &   -6.20  &    5.70  &   -3.59  &   17.04  &   13.20  &  -11.67  &   22.48  &   52.46  &   49.45  &   -0.85  &   60.10 \cr 
$\langle I_2A\rangle $  &    2.88  &   -9.70  &  -14.49  &  -27.06  &  -17.01  &  -13.01  &   17.12  &  -46.55  &  -32.44  &    6.94  &   70.26 \cr 
$\langle I_1I_2\rangle $  &    1.41  &   -2.79  &   22.62  &  -22.64  &   13.04  &  -43.72  &   76.79  &   -2.76  &    1.09  &    7.95  &  -30.27 \cr 
\end{tabular}
\end{table}
\begin{table}
\centering

\begin{tabular}{|| c | c | c | c | c | c | c | c | c | c | c ||}
$\delta h/h$ & $\delta \Omega _b/\Omega _b$ & $\delta \Omega _k$ & $\delta \Omega _\Lambda /\Omega _\Lambda $ & $\delta n_s/n_s$ & $\tau _{reion}$ & $\langle I_1I_1\rangle $ & $\langle I_2I_2\rangle $ & $\langle I_1A\rangle $ & $\langle I_2A\rangle $ & $\langle I_1I_2\rangle $ \cr
\hline
  12.71  &   20.88  &    3.10  &   10.62  &    5.52  &    2.08  &    4.85  &    9.47  &   23.18  &   26.61  &   12.35 \cr 
\end{tabular}
\end{table}
\centerline{\bf Table 6-M-TP}

\vfill\eject
\begin{table}
\centering

\begin{tabular}{|| c || c | c | c | c | c | c | c | c | c | c | c ||}
& 1 & 2 & 3 & 4 & 5 & 6 & 7 & 8 & 9 & 10 & 11 \cr
\hline
\% error &   0.03  &    0.23  &    0.28  &    0.92  &    1.53  &    2.11  &    2.74  &    5.71  &   16.09  &   34.79  &   46.10 \cr 
\hline
$\delta h/h$  &  -26.44  &   23.26  &   76.15  &    9.42  &  -26.21  &    2.62  &   12.98  &   -6.75  &   13.84  &   30.40  &   29.16 \cr 
$\delta \Omega _b/\Omega _b$  &    7.36  &   11.23  &   42.95  &    2.68  &  -18.94  &    0.11  &    3.36  &    9.58  &  -27.28  &  -61.58  &  -54.49 \cr 
$\delta \Omega _k$  &   87.19  &   -4.73  &   25.67  &   16.18  &    7.93  &  -28.72  &  -15.28  &   -5.85  &    3.36  &   14.15  &    9.37 \cr 
$\delta \Omega _\Lambda /\Omega _\Lambda $  &   40.44  &   20.44  &  -14.99  &  -26.46  &  -35.55  &   60.98  &   43.65  &    4.94  &    6.13  &    0.99  &    8.41 \cr 
$\delta n_s/n_s$  &   -1.24  &  -92.92  &   18.67  &   10.32  &  -12.86  &   23.14  &   14.05  &    2.96  &    0.59  &    1.26  &    0.10 \cr 
$\tau _{reion}$  &    0.00  &    2.75  &   -0.32  &   15.48  &   10.38  &   -2.01  &   16.25  &    1.36  &   84.49  &    3.19  &  -47.19 \cr 
$\langle I_1I_1\rangle $  &   -2.17  &    3.19  &  -29.51  &   52.85  &  -64.24  &   -9.04  &  -12.90  &  -43.20  &    2.36  &   -8.41  &    1.94 \cr 
$\langle I_2I_2\rangle $  &    0.09  &   12.04  &   -2.16  &   68.91  &   42.33  &   19.74  &   37.63  &   13.54  &   -9.90  &  -23.58  &   25.81 \cr 
$\langle I_1A\rangle $  &   -1.76  &   -3.22  &  -10.87  &   -2.55  &   -8.63  &  -51.21  &   70.25  &   -1.76  &  -27.35  &   30.48  &  -23.44 \cr 
$\langle I_2A\rangle $  &    0.55  &    7.80  &    5.74  &   18.89  &   22.50  &   42.60  &  -15.65  &  -31.77  &  -31.24  &   50.91  &  -49.05 \cr 
$\langle I_1I_2\rangle $  &    0.19  &   -4.99  &    9.88  &  -26.29  &   29.23  &   -1.06  &   21.37  &  -82.04  &    7.04  &  -30.37  &   13.04 \cr 
\end{tabular}
\end{table}
\begin{table}
\centering

\begin{tabular}{|| c | c | c | c | c | c | c | c | c | c | c ||}
$\delta h/h$ & $\delta \Omega _b/\Omega _b$ & $\delta \Omega _k$ & $\delta \Omega _\Lambda /\Omega _\Lambda $ & $\delta n_s/n_s$ & $\tau _{reion}$ & $\langle I_1I_1\rangle $ & $\langle I_2I_2\rangle $ & $\langle I_1A\rangle $ & $\langle I_2A\rangle $ & $\langle I_1I_2\rangle $ \cr
\hline
  17.26  &   33.31  &    6.62  &    4.43  &    0.85  &   25.68  &    4.12  &   14.63  &   15.92  &   29.24  &   13.10 \cr 
\end{tabular}
\end{table}
\centerline{\bf Table 6-P-T}

\vfill\eject
\begin{table}
\centering

\begin{tabular}{|| c || c | c | c | c | c | c | c | c | c | c | c ||}
& 1 & 2 & 3 & 4 & 5 & 6 & 7 & 8 & 9 & 10 & 11 \cr
\hline
\% error &   0.02  &    0.16  &    0.20  &    0.31  &    0.40  &    0.61  &    0.85  &    1.16  &    1.37  &    3.78  &    6.90 \cr 
\hline
$\delta h/h$  &  -25.66  &  -72.88  &  -31.23  &   22.24  &  -14.57  &  -16.36  &   -5.07  &   -1.03  &   -3.29  &  -44.77  &   -6.14 \cr 
$\delta \Omega _b/\Omega _b$  &    7.20  &  -39.06  &  -16.14  &   10.52  &   -6.66  &  -15.34  &  -14.59  &  -19.08  &    8.33  &   83.64  &   11.38 \cr 
$\delta \Omega _k$  &   88.09  &  -25.37  &    0.60  &   14.29  &   -0.62  &   11.94  &   24.35  &   19.70  &  -13.79  &   -8.63  &   -1.69 \cr 
$\delta \Omega _\Lambda /\Omega _\Lambda $  &   38.92  &   19.83  &  -18.19  &  -12.79  &  -10.34  &  -38.69  &  -53.44  &  -43.06  &   25.33  &  -25.13  &   -3.29 \cr 
$\delta n_s/n_s$  &   -0.41  &  -25.52  &   88.53  &   22.07  &   -0.21  &   -0.15  &  -18.42  &  -22.77  &   10.44  &   -6.52  &   -3.97 \cr 
$\tau _{reion}$  &   -0.03  &    3.77  &   -8.29  &   36.56  &   88.78  &  -22.80  &  -10.74  &    7.55  &   -0.43  &   -1.40  &   -1.86 \cr 
$\langle I_1I_1\rangle $  &   -3.01  &   33.14  &   -7.53  &   75.14  &  -32.47  &  -14.22  &   28.96  &   -4.23  &   30.74  &   -0.89  &   11.46 \cr 
$\langle I_2I_2\rangle $  &    0.02  &    3.99  &  -17.00  &   21.78  &    9.13  &   60.86  &   -9.15  &  -59.84  &  -33.04  &   -7.61  &   24.88 \cr 
$\langle I_1A\rangle $  &   -2.15  &   12.29  &    4.80  &    6.80  &   -8.96  &  -39.87  &   23.30  &  -30.01  &  -64.04  &    8.35  &  -49.91 \cr 
$\langle I_2A\rangle $  &    0.71  &   -6.51  &  -11.66  &   -2.48  &   11.77  &   35.00  &   16.55  &  -20.95  &   48.01  &    5.65  &  -73.50 \cr 
$\langle I_1I_2\rangle $  &    0.43  &  -12.35  &    4.75  &  -32.10  &   19.93  &  -24.84  &   64.18  &  -43.35  &   23.17  &   -9.44  &   34.01 \cr 
\end{tabular}
\end{table}
\begin{table}
\centering

\begin{tabular}{|| c | c | c | c | c | c | c | c | c | c | c ||}
$\delta h/h$ & $\delta \Omega _b/\Omega _b$ & $\delta \Omega _k$ & $\delta \Omega _\Lambda /\Omega _\Lambda $ & $\delta n_s/n_s$ & $\tau _{reion}$ & $\langle I_1I_1\rangle $ & $\langle I_2I_2\rangle $ & $\langle I_1A\rangle $ & $\langle I_2A\rangle $ & $\langle I_1I_2\rangle $ \cr
\hline
   1.76  &    3.27  &    0.51  &    1.26  &    0.54  &    0.44  &    0.97  &    1.97  &    3.60  &    5.13  &    2.51 \cr 
\end{tabular}
\end{table}
\centerline{\bf Table 6-P-TP}

\vfill\eject
\begin{table}
\centering

\begin{tabular}{|| c || c | c | c | c | c | c | c | c | c | c | c ||}
& 1 & 2 & 3 & 4 & 5 & 6 & 7 & 8 & 9 & 10 & 11 \cr
\hline
\% error &   0.14  &    0.39  &    0.87  &    1.65  &    2.42  &    6.18  &    6.82  &   19.15  &   51.30  &  140.60  &  451.42 \cr 
\hline
$\delta h/h$  &   17.65  &   69.39  &   20.75  &   29.55  &  -23.01  &  -12.27  &   15.26  &   -5.27  &    2.37  &   47.70  &   18.64 \cr 
$\delta \Omega _b/\Omega _b$  &  -12.10  &   35.97  &   14.08  &   12.08  &  -18.04  &   -7.20  &  -15.46  &  -39.29  &  -35.72  &  -51.67  &  -45.94 \cr 
$\delta \Omega _k$  &  -88.63  &    7.25  &   -1.08  &   16.78  &   -6.53  &   16.22  &   33.18  &    1.00  &   13.98  &   -3.32  &   13.97 \cr 
$\delta \Omega _\Lambda /\Omega _\Lambda $  &  -39.50  &    4.16  &    3.38  &  -45.98  &    0.71  &  -33.13  &  -48.09  &   -2.36  &  -23.27  &   47.41  &   -9.52 \cr 
$\delta n_s/n_s$  &   -1.43  &  -58.68  &   54.27  &   36.48  &  -35.61  &   -9.33  &    2.45  &  -10.50  &  -16.97  &   21.71  &   -6.96 \cr 
$\tau _{reion}$  &    0.57  &    3.05  &  -15.26  &   10.03  &   11.54  &   18.43  &   28.36  &   51.77  &  -69.42  &   15.96  &  -25.01 \cr 
$\langle I_1I_1\rangle $  &    7.16  &  -13.76  &  -46.94  &  -23.99  &  -44.80  &  -38.78  &   51.14  &  -23.55  &   -0.73  &    8.10  &  -15.29 \cr 
$\langle I_2I_2\rangle $  &   -3.74  &  -10.37  &  -57.42  &   56.15  &    7.83  &   14.19  &  -33.69  &  -25.66  &    6.80  &   29.85  &  -20.85 \cr 
$\langle I_1A\rangle $  &    6.16  &   -5.62  &   -2.72  &  -22.36  &    5.93  &   47.19  &    9.31  &  -57.98  &  -41.52  &   11.82  &   42.99 \cr 
$\langle I_2A\rangle $  &   -4.58  &   -5.91  &  -12.90  &   30.41  &   28.32  &  -61.39  &   -2.27  &   -2.00  &  -31.33  &  -23.47  &   52.35 \cr 
$\langle I_1I_2\rangle $  &    0.33  &   -2.44  &   22.52  &   -0.28  &   69.25  &  -17.59  &   38.23  &  -32.44  &   11.01  &   20.55  &  -36.40 \cr 
\end{tabular}
\end{table}
\begin{table}
\centering

\begin{tabular}{|| c | c | c | c | c | c | c | c | c | c | c ||}
$\delta h/h$ & $\delta \Omega _b/\Omega _b$ & $\delta \Omega _k$ & $\delta \Omega _\Lambda /\Omega _\Lambda $ & $\delta n_s/n_s$ & $\tau _{reion}$ & $\langle I_1I_1\rangle $ & $\langle I_2I_2\rangle $ & $\langle I_1A\rangle $ & $\langle I_2A\rangle $ & $\langle I_1I_2\rangle $ \cr
\hline
 107.64  &  220.62  &   63.70  &   80.32  &   44.74  &  120.93  &   70.24  &  103.29  &  196.28  &  239.18  &  167.06 \cr 
\end{tabular}
\end{table}
\centerline{\bf Table 7-M-T}

\vfill\eject
\begin{table}
\centering

\begin{tabular}{|| c || c | c | c | c | c | c | c | c | c | c | c ||}
& 1 & 2 & 3 & 4 & 5 & 6 & 7 & 8 & 9 & 10 & 11 \cr
\hline
\% error &   0.14  &    0.39  &    0.85  &    1.53  &    1.77  &    2.29  &    4.98  &    5.81  &   10.55  &   26.75  &   53.08 \cr 
\hline
$\delta h/h$  &   17.45  &   69.55  &   20.11  &  -24.73  &  -17.72  &   23.02  &   19.62  &    3.76  &   14.00  &   34.83  &   34.49 \cr 
$\delta \Omega _b/\Omega _b$  &  -12.14  &   35.97  &   13.57  &  -10.19  &   -6.64  &   16.63  &   -9.54  &   22.44  &   38.00  &  -60.46  &  -47.49 \cr 
$\delta \Omega _k$  &  -88.74  &    7.18  &   -1.21  &  -13.91  &   -8.85  &    7.90  &   24.06  &  -31.86  &  -11.14  &    1.32  &    4.28 \cr 
$\delta \Omega _\Lambda /\Omega _\Lambda $  &  -39.31  &    3.87  &    3.61  &   41.26  &   19.73  &   -1.57  &  -31.95  &   55.98  &   29.60  &   28.87  &   21.65 \cr 
$\delta n_s/n_s$  &   -1.37  &  -58.49  &   52.49  &  -34.86  &  -13.30  &   38.96  &    4.24  &   13.48  &   23.97  &    8.11  &    8.41 \cr 
$\tau _{reion}$  &    0.70  &    3.48  &  -25.12  &  -38.54  &   81.65  &   34.13  &   -5.09  &   -0.19  &   -2.64  &    0.47  &    2.32 \cr 
$\langle I_1I_1\rangle $  &    7.18  &  -13.86  &  -46.38  &   31.74  &  -11.73  &   42.58  &   61.68  &   16.87  &   20.24  &    5.40  &  -10.94 \cr 
$\langle I_2I_2\rangle $  &   -3.85  &  -10.30  &  -56.78  &  -47.26  &  -32.63  &  -17.37  &  -30.27  &   -0.69  &   40.27  &   22.22  &   -4.73 \cr 
$\langle I_1A\rangle $  &    6.19  &   -5.68  &   -2.56  &   21.15  &   10.57  &   -2.65  &   -2.54  &  -45.64  &   52.46  &  -37.83  &   55.53 \cr 
$\langle I_2A\rangle $  &   -4.58  &   -5.76  &  -12.79  &  -30.07  &   -5.13  &  -29.93  &   29.06  &   52.74  &  -20.87  &  -42.20  &   45.97 \cr 
$\langle I_1I_2\rangle $  &    0.41  &   -2.29  &   21.56  &   -9.76  &   31.46  &  -58.55  &   48.05  &   -1.51  &   39.48  &   22.22  &  -25.55 \cr 
\end{tabular}
\end{table}
\begin{table}
\centering

\begin{tabular}{|| c | c | c | c | c | c | c | c | c | c | c ||}
$\delta h/h$ & $\delta \Omega _b/\Omega _b$ & $\delta \Omega _k$ & $\delta \Omega _\Lambda /\Omega _\Lambda $ & $\delta n_s/n_s$ & $\tau _{reion}$ & $\langle I_1I_1\rangle $ & $\langle I_2I_2\rangle $ & $\langle I_1A\rangle $ & $\langle I_2A\rangle $ & $\langle I_1I_2\rangle $ \cr
\hline
  20.64  &   30.25  &    3.41  &   14.67  &    5.75  &    2.18  &    7.22  &    7.95  &   31.77  &   27.20  &   15.64 \cr 
\end{tabular}%
\end{table}
\centerline{\bf Table 7-P-TP}

\vfill\eject
\begin{figure}
\centerline{\psfig{file=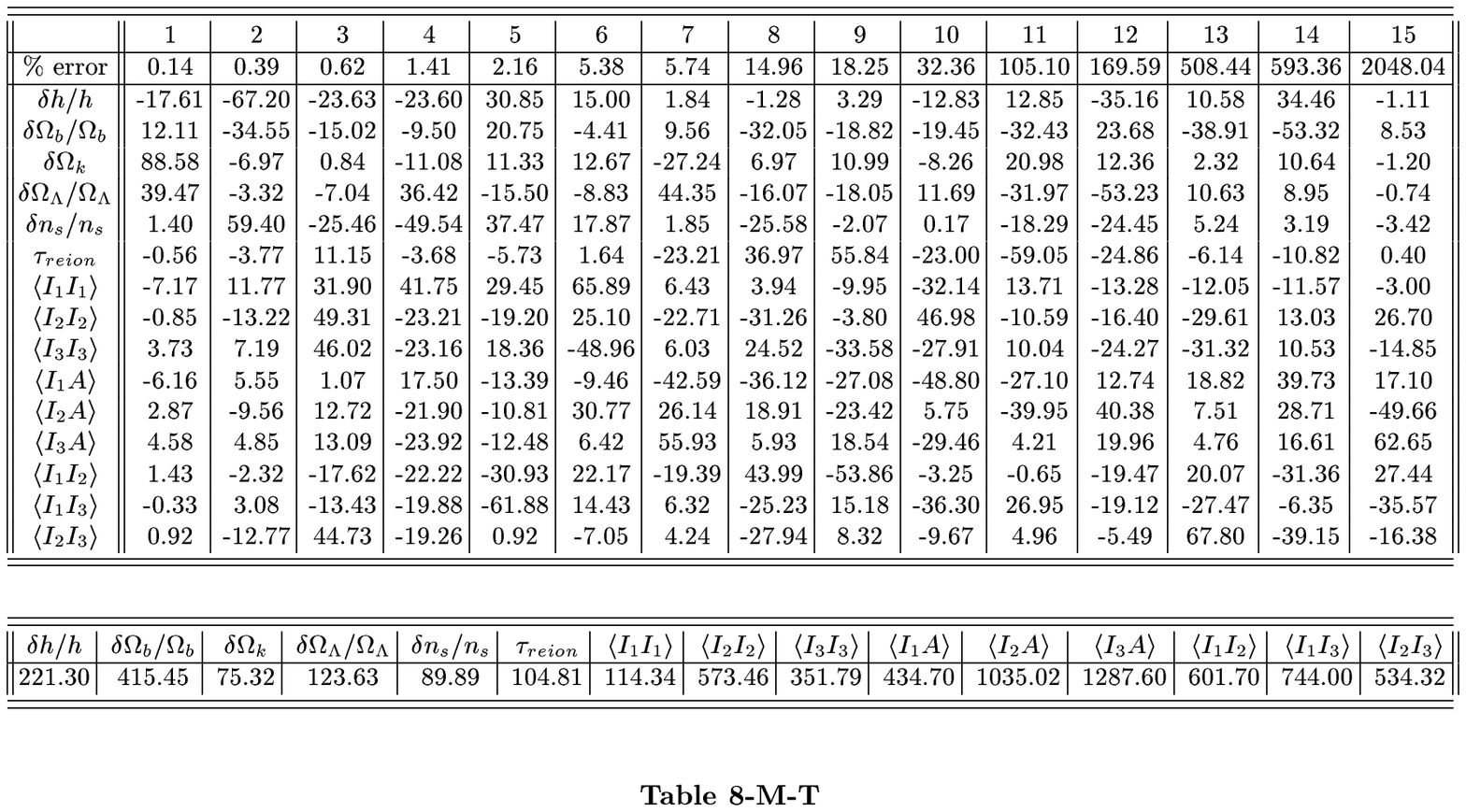,width=8.in}}
\end{figure}
\vskip -6in
\begin{figure}
\centerline{\psfig{file=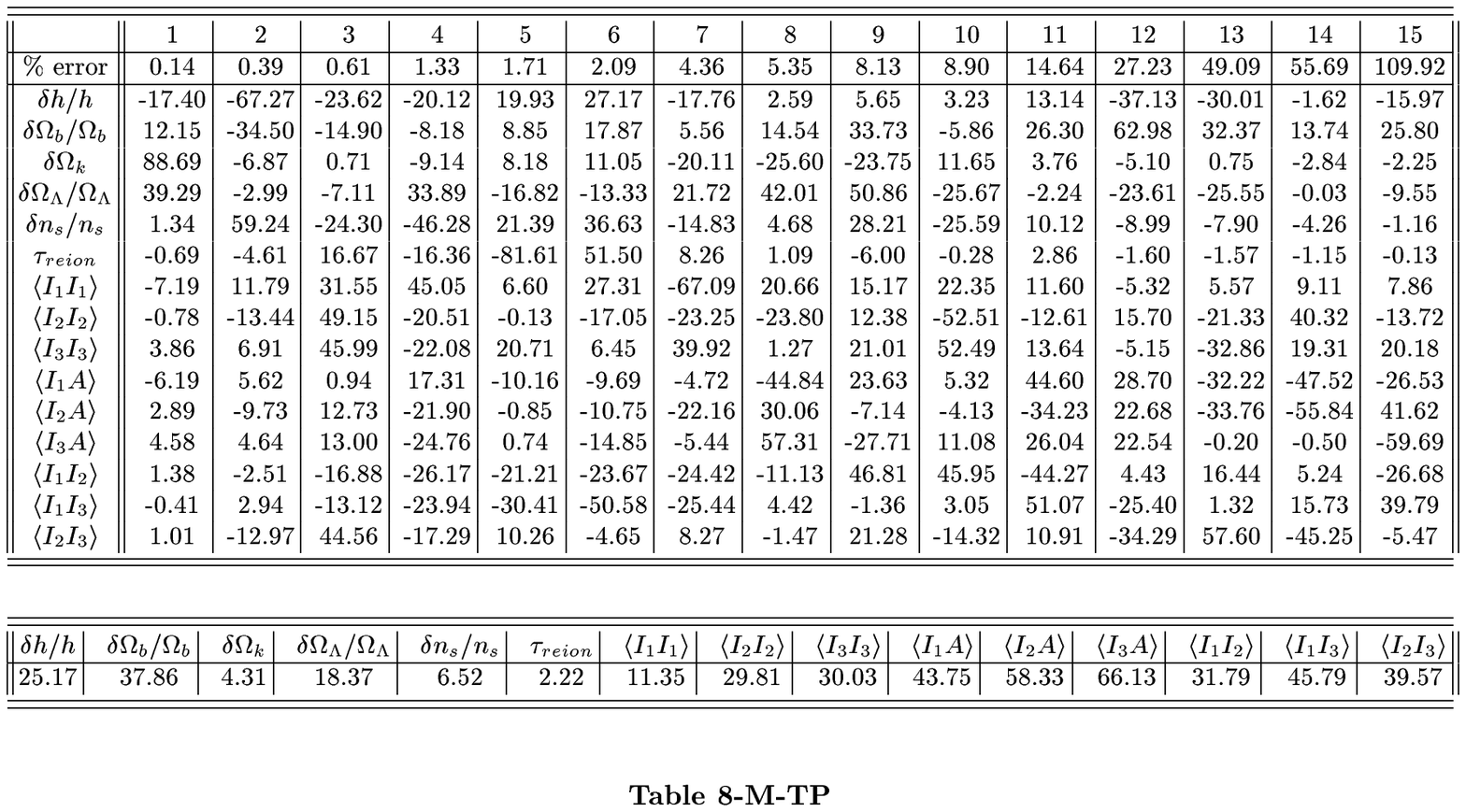,width=8.in}}
\end{figure}
\begin{figure}
\centerline{\psfig{file=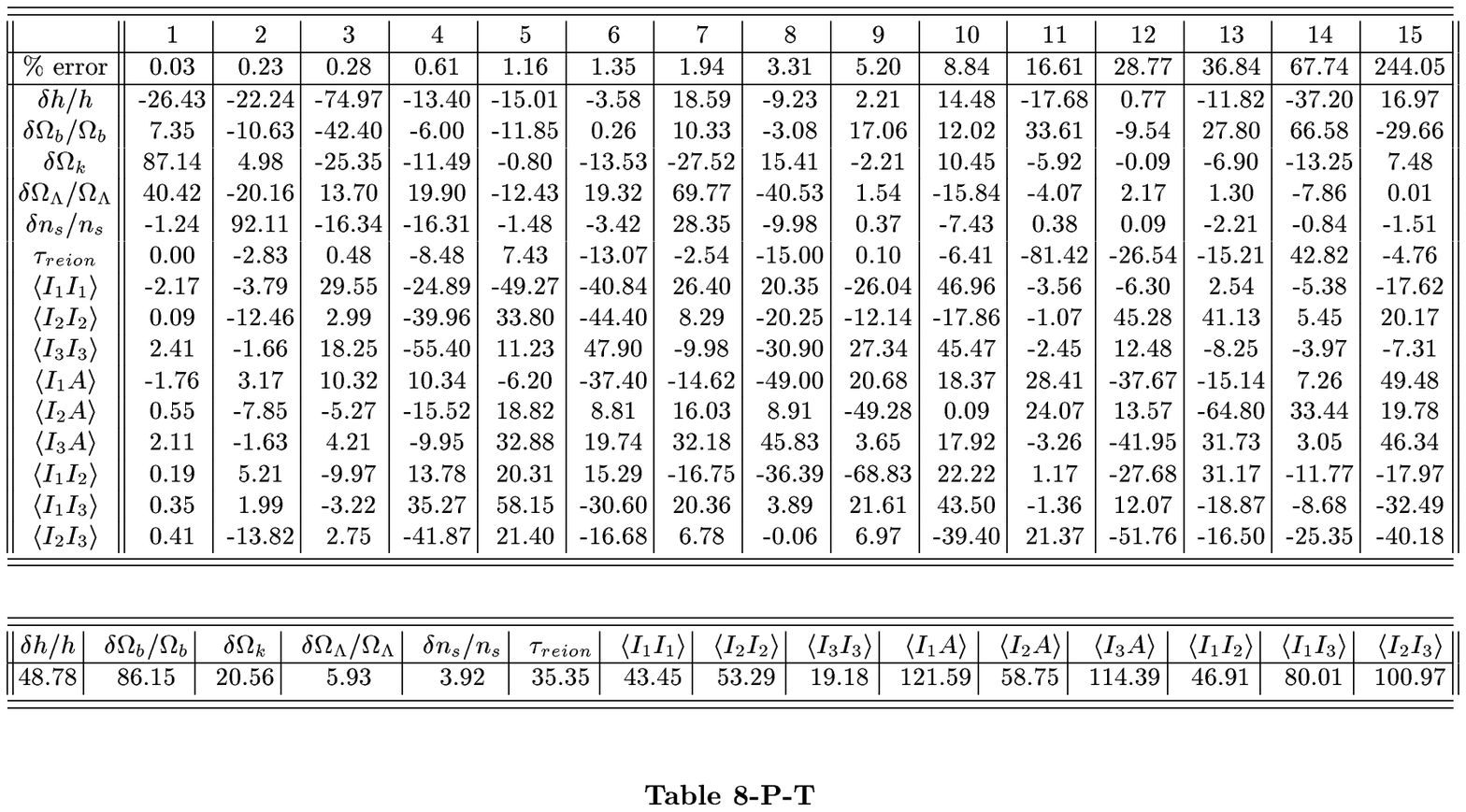,width=8.in}}
\end{figure}
\begin{figure}
\centerline{\psfig{file=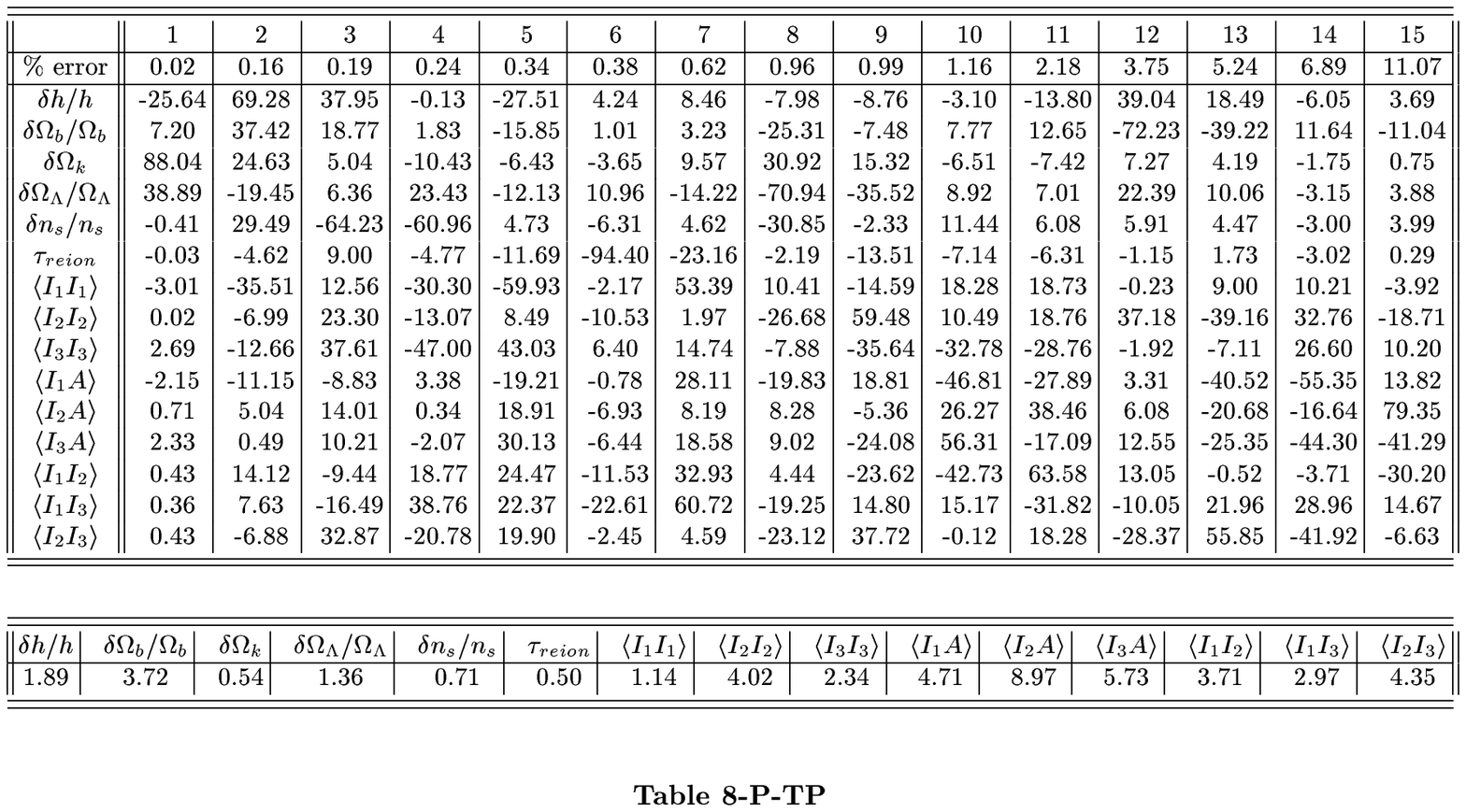,width=8.in}}
\end{figure}

\begin{thebibliography}{99}

\bibitem{cobe}Bennet, C. {\it et al.}, ``4-Year COBE DMR Cosmic Microwave Background
Observations: Maps and Basic Results," 
Ap.J. 464 (1996) L1 (astro-ph/9601067) 

\bibitem{toco}  Miller, A. {\it et al.}, 
``A Measurement of the Angular Power Spectrum of the CMB from l = 100 to 400'',
astro-ph/9906421, Ap. J. 524 (1999) L1.

\bibitem{boom} P. de Bernardis {\it et al.}, ``A flat universe from high-resolution 
maps of the cosmic microwave background radiation," Nature, 404, 955, (2000). 
       
\bibitem{maxima}
Hanany, S. {\it et al.}, "MAXIMA-1: A Measurement of the Cosmic 
Microwave Background Anisotropy on angular scales of 10 arcminutes to 
5 degrees," (astro-ph/0005123) (Submitted to Astrophys. J. Lett.)

\bibitem{mapp}http://map.gsfc.nasa.gov/

\bibitem{plnck}http://astro.estec.esa.nl/SA-general/Projects/Planck/

\bibitem{paramest}See, for example:
Bond, J.R., Efstathiou, G. \& Tegmark, M. 1997,
``Forecasting Cosmic Parameter Errors from
Microwave Background Anisotropy Experiments,"
M.N.R.A.S., 291, L33; Knox, L. 1995, Phys. Rev. D, 53 4307;
Jungman, G., Kamionkowski, M., Kosowsky, A.,
\& Spergel, D. 1996, Phys. Rev. Lett., 76, 1007;
Zaldarriaga, M., Spergel, D.N. \& Seljak, U.
1997, ``Microwave Background Constraints on Cosmological Parameter,"
Ap.J. 488, 1. A recent reference is W. Hu, M. Fukugita, M. Zaldarriaga and
M. Tegmark, ``CMB Observables and Their Cosmological Implications'',
astro-ph/0006436 (2000).


\bibitem{infpert}Hawking, S.W., 1982, ``The Development of Irregularities
in a Single Bubble Inflationary Universe," Phys. Lett., 115B, 295;
Starobinsky, A.A., 1982, ``Dynamics of Phase Transition in the New
Inflationary Scenario and the Generation of Perturbations,"
Phys. Lett., 117B, 125; Guth, A. \& Pi, S.Y., 1982,
``Fluctuations in the New Inflationary Universe,"
Phys. Rev. Lett., 49, 1110;
Bardeen, J., Steinhardt, P., \& Turner, M. 1983,
``Spontaneous Generation of Almost Scale-Free Density
Perturbations in an Inflationary Universe,"
Phys. Rev. D., 28, 679

\bibitem{liddle}A. Liddle and D. Lyth, {\it Cosmological Inflation
and Large-Scale Structure,}
(Cambridge: Cambridge University Press, 2000)

\bibitem{nongauss}See, for example:
Salopek, D.S., 1992,``Cold-Dark Matter Cosmology with
Non-Gaussian Fluctuations from Inflation," Phys. Rev., D45,
1139;
Kofman , L., \&  Linde, A. 1987, ``Generation of Density
Perturbations in Inflationary Cosmology," Nucl. Phys., B282,
555;
Vishniac, E., Olive, K. \& Seckel, D. 1987, ``Cosmic Strings and
Inflation," Nucl. Phys. B, 289, 717);
Kofman, L., Linde A., \& Einasto, J. 1987 ``Cosmic Bubbles as Remnants
from Inflation," Nature, 326, 48;
Hodges, H. \&Primack, J. 1991, ``Strings, Textures and Inflation,"
Phys. Rev. D, 43, 3155;
Linde, A. 1992, ``Strings, Textures, Inflation, and Spectrum Bending,"
Phys. Lett 284B, 215 (1992);
Basu, R., Guth, A. \& Vilenkin, A. 1991, ``Quantum
Creation of Topological Defects During Inflation," Phys. Rev. D, 44,
340;
Kofman, L. 1986, ``What Initial Perturbations May Be Generated in Inflationary
Cosmological Models," Phys. Lett., 173B, 400;
Bucher, M. \& Zhu, Y. 1997,, ``Non-Gaussian
Isocurvature Perturbations From Goldstone Modes
Generated During Inflation," Phys.Rev. D, 55, 7415;
and references therein.

\bibitem{bmt}
Bucher, M., Moodley, K. \& Turok, N., 1999, The General 
Primordial Cosmic Perturbation," (astro-ph/9904231)
(to appear in Phys. Rev. D)

\bibitem{pjepa}Peebles, P.J.E. 1987, ``Origin of the 
Large Scale Peculiar Velocity Field: A Minimal 
Isocurvature Model," Nature 327, 210;
Peebles, P.J.E. 1987, ``Cosmic Background Temperature
Anisotropy in a Minimal Isocurvature Model for Galaxy
Formation," Ap.J. 315, L73

\bibitem{bondb}Bond, J.R. \& Efstathiou, G. 1987,
``Microwave Anisotropy Constraints on Isocurvature Baryon Models,"
M.N.R.A.S., 22, 33

\bibitem{cen}Cen, R., Ostriker, J.P., \& Peebles, P.J.E. 1993,
``A Hydrodynamic Approach to Cosmology: The Primeval
Baryon Isocurvature Model," Ap.J., 415, 423

\bibitem{bond}Bond, J.R. \& Efstathiou, G. 1986, 
``Isocurvature cold dark matter fluctuations,"
M.N.R.A.S., 218, 103

\bibitem{nnewb}Rebhan, A. \& Schwarz, D. 1994, ``Kinetic versus
Thermal Field Theory Approach to Cosmological Perturbations,"
Phys. Rev., D50, 2541

\bibitem{nnewc}Challinor, A. \& Lasenby, A. 1999, ``Cosmic
Microwave Background Anisotropy in the Cold Dark Matter Model:
A Covariant and Gauge Invariant Approach," Ap. J., 513, 1

\bibitem{stompor}Stompor, R., Banday, A.J., \& Gorski, K.M.~1996,
``Flat Dark Matter-Dominated Models with Hybrid Adiabatic Plus
Isocurvature Initial Conditions," Ap.J., 463, 8

\bibitem{gouda}Gouda, N. \& Sugiyama, N.~1992, ``Surviving Cosmological
Models After the Discovery of Large-Angle Anisotropies in the
Microwave Background," Ap.J., 395L, 59

\bibitem{yo}Gorski, K.~\& Silk, J. 1989, ``Large Scale Microwave Background
Anisotropies in Isocurvature Baryon Open Universe Models," Ap.J., 346, 1

\bibitem{gor}Yokoyama, J. \& Suto, Y. ``Baryon Isocurvature Scenario in Inflationary
Cosmology: A Particle Physics Model and its Astrophysical Implications,"
Ap.J., 379, 427

\bibitem{sugi}Suginohara, T., \& Suto, Y.~1992, ``Large Scale Structure in Isocurvature
Baryon Models," Ap.J., 387, 431

\bibitem{chiba}Chiba, T., Sugiyama, N., \& Suto Y.~1994, ``Microwave Background Anisotropies
in Primeval Baryon Isocurvature Models: Constraints on the Cosmological Parameters,"
Ap.J., 429, 427

\bibitem{hu}Hu, W., Bunn, E., \& Sugiyama, N. 1995, ``COBE 
Constraints on Baryon Isocurvature Models," Ap.J., 447, 59

\bibitem{nnewa}Sugiyama, N., Sasaki, M., \& Tomita, K. 1989, 
``Cosmic Microwave Background Anisotropies and Large-Scale 
Velocity Fields in Isocurvature Hot Dark Matter Models,"
Ap. J. 338, L45
 
\bibitem{nnewd}Yokoyama, J. 1994, ``Formation of Baryon Number
Fluctuation in Supersymmetric Inflationary Cosmology," 
Astroparticle Physics 2, 291

\bibitem{enqvist}
K. Enqvist and H. Kurki-Suonio,
``Constraining Isocurvature Fluctuations with Planck Surveyor,"
Phys. Rev. D61, 43002 (2000) (astro-ph/9907221);
K. Enqvist and H. Kurki-Suonio,,
``Limits on Isocurvature Fluctuations From Boomerang and Maxima,"
(astro-ph/0006429)

\bibitem{pierpaoli}
E. Pierpaoli, J. Garci\'a-Bellido and S. Borgani,
``Microwave Background Anisotropies and Large Scale Structure Constraints
on Isocurvature Modes in a Two Field Model of Inflation,"
JHEP 9910, 15 (1999) (hep-ph/9909420)

\bibitem{lindeb}
A. Linde and S. Mukhanov, 
``Non-Gaussian Isocurvature Perturbations from Inflation,"
Phys. Rev. D56, 535 (1997) (astro-ph/9610219)

\bibitem{langl}
D. Langlois, ``Correlated Adiabatic and Isocurvature Perturbations from
Double Inflation," Phys. Rev. D59, 123512, (1999) (astro-ph/9906080)

\bibitem{langlr}D. Langlois and A. Riazuelo,
``Correlated mixtures of adiabatic and isocurvature
cosmological perturbations," (astro-ph/9912497)

\bibitem{polarski}D. Polarski and A.A. Starobinskii,
``Isocurvature Perturbations in Multiple Inflationary Models,"
Phys. Rev. D50, 6123 (1994)

\bibitem{kendall}
A. Stuart and J.K. Ord, {\it Kendall's Advanced Theory of Statistics,}
(Oxford: Oxford University Press, 1991). 

\bibitem{wang}Y. Wang, D. Spergel, \& M. Strauss, ``Cosmology in the
Next Millenium: Combining MAP and SDSS Data to Constrain Inflationary
Models," Ap.J. 498, 1 (1998) (astro-ph/9802231) 

\bibitem{peenew}
Peebles, P.J.E. 1999, ``An Isocurvature Cold Dark Matter Cosmogony. 
I. A Worked Example of Evolution through Inflation," Ap.J., 510, 523;
Peebles, P.J.E. 1999, ``An Isocurvature Cold Dark Matter Cosmogony. 
II. Observational Tests," Ap.J. 510, 531 

\end{thebibliography}
\end{document}